\documentclass[%
reprint,
superscriptaddress,
nofootinbib,
amsmath,amssymb,
aps,
]{revtex4-2}

\usepackage[dvipdfmx]{graphicx}
\usepackage{dcolumn}
\usepackage{bm}

\begin{document}

\preprint{APS/123-QED}

\title{
Constraining self-interactions of a massive scalar field using scalar gravitational waves from stellar core collapse
}

\author{Naomichi Asakawa}
\affiliation{%
Department of Physics, Toho University, Funabashi, Chiba 274-8510, Japan
}%
\author{Yuichiro Sekiguchi}%
\affiliation{%
Department of Physics, Toho University, Funabashi, Chiba 274-8510, Japan
}%
\affiliation{%
Center for Gravitational Physics and Quantum Information,
Yukawa Institute for Theoretical Physics, Kyoto University, Kyoto 606-8502, Japan
}%


%

\date{\today}

\begin{abstract}

We perform a comprehensive numerical study of gravitational waves from stellar core collapse
in the massive scalar-tensor theory with the cubic and quartic self-interactions of the scalar field.
We investigate the dependence of gravitational waves on the self-interaction as well as the mass of the scalar field and the conformal factor.
We find that gravitational-wave spectra show a systematic difference between the cubic and quartic self-interactions.
We also find that this systematic difference is insensitive to the mass of the scalar field and the conformal factor.
Our results indicate that the type of the self-interaction could be constrained by observations of gravitational waves using the future-planned detectors.

\end{abstract}

\maketitle


\section{
INTRODUCTION
\label{sec:intro}
}
General relativity (GR) has passed a number of experimental
tests~\cite{psaltis2008probes,will2014confrontation,berti2015testing}
and now is considered as the standard theory of gravity.
However, there are several difficulties to be overcome in GR, such as
the accelerating expansion of the Universe and the renormalization of gravitational field.
The former is usually explained in the GR framework by introducing dark energy,
but it could be explained by a modification of gravity~\cite{li2011dark}.
As for the latter, the nonrenormalizability of GR implies
that GR is the effective theory at the low energy regime~\cite{burgess2004quantum}.
These difficulties motivate us to explore modifications of GR.

Scalar-tensor (ST) theories~\cite{damour1992tensor,fujii2003scalar}
are ones of the well-motivated classes of modified gravity.
In ST theories, a scalar field can cause the inflation~\cite{la1989extended}.
Furthermore, the so-called $f(R)$ gravity~\cite{sotiriou2010f}, which can be recast into
ST theories, can explain the accelerating expansion of the Universe~\cite{nojiri2003modified}.
In addition, the superstring theory, which is a candidate of the renormalizable theory of gravity,
induces a coupling of scalar fields with gravitation as in ST theories in the low energy
limit~\cite{callan1985strings}.
Therefore, ST theories are promising theories of modified gravity.

From a phenomenological viewpoint, ST theories are also interesting because they induce a nonperturbative phenomenon so-called spontaneous scalarization~\cite{damour1993nonperturbative,damour1996tensor},
in which neutron stars have a large scalar charge due to tachyonic instability of the scalar field.
Because neutron stars exhibiting a spontaneous scalarization are considered to be energetically favorable,
when the spontaneous scalarization occurs,
a new branch of stationary families of neutron stars emerges~\cite{damour1993nonperturbative}.
The structure of a neutron star in ST theories has been studied in the case of
the static~\cite{ramazanouglu2016spontaneous,morisaki2017spontaneous,rosca2020structure},
slowly rotating~\cite{staykov2018static,yazadjiev2016slowly},
and rapidly rotating cases~\cite{doneva2013rapidly}.
These studies show that
the spontaneous scalarization has a considerable effect on the structure of neutron stars.
In addition, a new possibility of the accretion induced descalarization of a neutron star has been proposed in Ref.~\cite{PhysRevLett.129.121104}.

The spontaneous scalarization could occur with appropriate parameters of ST theories.
The parameters in ST theories are constrained by several observations, such as
the Cassini satellite mission~\cite{bertotti2003test}
and the orbital decay of binary pulsars~\cite{antoniadis2013massive,freire2012relativistic}.
For ST theories with a massless scalar field, the range of parameters which allows the spontaneous
scalarization, is quite narrow
(see Refs.~\cite{novak1998neutron,doneva2013rapidly} for the condition of the parameter for the spontaneous scalarization
and Refs.~\cite{will2014confrontation,berti2015testing} for the constraint on the parameter).
For ST theories with a massive scalar field, on the other hand,
a broad region of parameters remains unconstrained for the scalar-field mass
$\mu \gtrsim 10^{-15}~\mathrm{eV}$~\cite{ramazanouglu2016spontaneous}.

It is known that ST theories allow scalar modes of gravitational waves (GWs) to be emitted~\cite{will2014confrontation}.
Hence, it is in principle possible to test ST theories by observations of GWs, and
detection of scalar modes can provide smoking gun evidence of the break of GR.
For the detection of scalar modes, four GW detectors are necessary~\cite{chatziioannou2012model}
when GW polarizations consist of two tensor modes
and two scalar modes.
\footnote{
When vector modes exist, six GW detectors are necessary to probe all of the polarizations of gravitational waves~\cite{chatziioannou2012model}.
}

Advanced LIGO~\cite{aasi2015advanced},
Advanced VIRGO~\cite{acernese2014advanced},
and KAGRA~\cite{somiya2012detector} are in operation, and GW observation O4 is running.
LIGO-India is planned to participate in the observation network from the late 2020s~\cite{abbott2020prospects},
and then it becomes possible to test ST theories by GWs.
Furthermore, third-generation GW detectors, such as
the Einstein Telescope~\cite{punturo2010einstein}
and Cosmic Explorer~\cite{abbott2017exploring},
are planned in the future.
Note that the amplitude of scalar modes is enhanced by the spontaneous scalarization~\cite{rosca2020core},
and scalar-mode GWs from strongly scalarized neutron stars could be detectable.

Gravitational collapse of a massive stellar core is one of the promising sources of GWs.
Massive stars with a zero-age-main-sequence (ZAMS) mass $M_\mathrm{ZAMS} \gtrsim 8M_{\odot}$
would undergo gravitational collapse at the end point of their evolution
and leave a neutron star or a black hole.
Stellar core collapse in the massive ST theory was studied in spherical symmetry for
the no self-interaction~\cite{sperhake2017long,rosca2020core},
the quartic self-interaction~\cite{rosca2019inverse,cheong2019numerical},
and the cubic self-interaction~\cite{huang2021scalar},
and basic features of scalar GWs are clarified as follows.
Scalar GWs become dispersive in their propagation, and as a consequence,
the observed scalar GW signals are inverse chirp,
quasimonochromatic,
and long-lived ones~\cite{sperhake2017long,rosca2020core}.
The amplitude of scalar GWs decreases as one increases the scalar-field mass~\cite{rosca2020core}.
The self-interactions of the scalar field suppress the scalar GW signals
in a frequency-dependent manner~\cite{cheong2019numerical,huang2021scalar}.
This suppression indicates that
if observed scalar GW signals are suppressed in such a way,
we may be able to refer to the existence of self-interactions~\cite{cheong2019numerical}.

The properties of scalar GWs from stellar core collapse, however, have not been studied
in detail, systematically changing the scalar-field mass and strength of self-interactions simultaneously.
In particular, the observed spectra of scalar GWs for the quartic self-interaction have not been clarified yet.
Thus, we performed a comprehensive numerical study of scalar GWs from collapse of a massive stellar
core in the massive ST theory, both with the cubic and quartic self-interactions using our newly developed code.
We systematically investigate the dependence of scalar GWs on the type and parameters of self-interactions as well as the scalar-field mass and the conformal factor.
As a result, we find that the type of the self-interactions can be constrained by observations of scalar GWs.

The paper is organized as follows.
In Sec.~\ref{sec:formalism}, formulation and basic equations in the massive ST theory with spherical symmetry are presented.
In Sec.~\ref{sec:SGW}, we outline the scalar GWs propagation.
In Sec.~\ref{sec:Setup}, we describe our numerical setup,
and in Sec.~\ref{sec:Results}, we present results of numerical simulations.
Finally, in Sec.~\ref{sec:Conclusion}, we present the summary.
Throughout the paper,
we use geometrical units $c=G=1$, where $c$ is the speed of light and $G$ is the gravitational constant, respectively.

\section{
FORMULATION
\label{sec:formalism}
}
In this section, we summarize formulation of the equations of motion for the metric, scalar field, and matter, following~\cite{gerosa2016numerical,rosca2020core}.

\subsection{Metric and scalar field}

The action of the ST theory in the Jordan frame can be written as~\cite{berti2015testing,salgado2006cauchy}
\begin{eqnarray}
  \label{Eq:Action_Jordan-frame}
  S_{J}
  &=&
  \int
  d^{4}x~
  \sqrt{-g}
  \biggr(
  \frac{1}{16\pi}
  F(\phi)R
  -\frac{1}{2}
  g^{\mu\nu}\partial_\mu\phi\partial_\nu\phi
  -U(\phi)
  \biggr)
  \nonumber\\
  &&
  +S_\mathrm{mat}[g_{\mu\nu}]
  ~,
\end{eqnarray}
where $g_{\mu\nu}$ is the metric in the Jordan frame,
$g$ is its determinant,
$R$ is the Ricci scalar with respect to the metric $g_{\mu\nu}$,
and $\phi$ is the scalar field, respectively.
$U(\phi)$ and $F(\phi)$ are
the potential and the conformal factor associated with the scalar field $\phi$.
$S_\mathrm{mat}$ is the action of the matter fields.

The action Eq.~(\ref{Eq:Action_Jordan-frame}) can be written in the different form,
in the so-called Einstein frame,
where the nonminimal coupling between the scalar field and the metric is removed.
This can be done by performing a conformal transformation,
\begin{equation}
  \tilde{g}_{\mu\nu}:=F(\phi)g_{\mu\nu}
  ~,
\end{equation}
and a redefinition of the scalar field by
\begin{eqnarray}
  \frac{\partial\varphi}{\partial\phi}
  :=
  \sqrt{
  \frac{3}{4}
  \biggr(\frac{\partial}{\partial\phi}\ln F\biggr)^{2}
  +\frac{4\pi}{F}
  }
  ~.
\end{eqnarray}
The resulting action in the Einstein frame is given by
\begin{eqnarray}
  \label{Eq:Action_Einstein-frame}
  S_{E}
  &=&
  \int
  d^{4}x~
  \frac{\sqrt{-\tilde{g}}}{16\pi}
  \biggr(
  \tilde{R}
  -2\tilde{g}^{\mu\nu}\partial_\mu\varphi\partial_\nu\varphi
  -4V(\varphi)
  \biggr)
  \nonumber\\&&
  +S_\mathrm{mat}[\tilde{g}_{\mu\nu}/F]
  ~,
\end{eqnarray}
where $\tilde{g}$ is the determinant of $\tilde{g}_{\mu\nu}$,
$\tilde{R}$ is the Ricci scalar with respect to $\tilde{g}_{\mu\nu}$,
and $V(\varphi):=4\pi U(\phi)/F(\phi)^{2}$, respectively.

The variation of the action Eq.~(\ref{Eq:Action_Einstein-frame}) gives the equations of motion in the Einstein frame as
\begin{eqnarray}
  &&\tilde{G}_{\mu\nu}
  =
  2\partial_{\mu}\varphi\partial_{\nu}\varphi
  -\tilde{g}_{\mu\nu}
  \partial^{\lambda}\varphi\partial_{\lambda}\varphi
  -2\tilde{g}_{\mu\nu}V
  +8\pi\tilde{T}_{\mu\nu}
  ~,~~~
  \label{Eq:Equations-Of-Motion_metric}
  \\&&
  \tilde{\nabla}^{\lambda}
  \tilde{\nabla}_{\lambda}
  \varphi
  =
  2\pi\tilde{T}
  \partial_{\varphi}\ln F
  +\partial_{\varphi}V
  ~,
  \label{Eq:Equations-Of-Motion_scalar}
\end{eqnarray}
where $\tilde{G}_{\mu\nu}$ is the Einstein tensor with respect to $\tilde{g}_{\mu\nu}$,
$\tilde{\nabla}_{\mu}$ is the covariant derivative associated with $\tilde{g}_{\mu\nu}$,
and $\partial_{\varphi}$ is the partial derivative with respect to the scalar field $\varphi$.
$T_{\mu\nu}=-2(-g)^{-1/2}\delta S_\mathrm{mat}/\delta g^{\mu\nu}$ and
$\tilde{T}_{\mu\nu}=T_{\mu\nu}/F$ are the energy-momentum tensor of matter fields in the Jordan frame
and Einstein frame, respectively, and $\tilde{T}=\tilde{g}^{\mu\nu}\tilde{T}_{\mu\nu}$.

As in the previous studies~
\cite{gerosa2016numerical,sperhake2017long,rosca2019inverse,rosca2020core,cheong2019numerical,huang2021scalar},
we consider stellar core collapse in spherical symmetry.
Then, the line element in the Einstein frame is given by
\begin{eqnarray}
  \label{Eq:Line-Element}
  d\tilde{s}^{2}
  =
  \tilde{g}_{\mu\nu}
  dx^{\mu}
  dx^{\nu}
  =
  -F\alpha^{2}dt^{2}
  +FX^{2}dr^{2}
  +r^{2}d\Omega^{2}
  ~.
\end{eqnarray}
Here, $\alpha$ and $X$ can be written as
\begin{eqnarray}
  \label{Eq:Metric-function}
  \sqrt{F}\alpha
  =
  \mathrm{exp}(\Phi)
  ~,~~
  \sqrt{F}X
  =
  \biggr(
  1-\frac{2m}{r}
  \biggr)^{-1/2}
  ~,
\end{eqnarray}
where $\Phi$ is the metric potential and $m$ is the mass function~\cite{gerosa2016numerical}.
Then, the equations of motion for the metric Eq.~(\ref{Eq:Equations-Of-Motion_metric}) are given by
\begin{eqnarray}
  \label{Eq:EOM_Spherical_metric_1}
  &&\partial_{r}\Phi
  =
  FX^{2}
  \biggr[
  \frac{m}{r^{2}}
  +4\pi r
  \tilde{g}^{rr}
  \tilde{T}_{rr}
  +\frac{r}{2F}
  (\eta^{2}+\psi^{2})
  \biggr]
  \nonumber\\
  &&~~~~~
  -FX^{2}rV
  ~,
  \\
  \label{Eq:EOM_Spherical_metric_2}
  &&\partial_{r}m
  =
  -4\pi r^{2}
  \tilde{g}^{tt}
  \tilde{T}_{tt}
  +\frac{r^{2}}{2F}
  (\eta^{2}+\psi^{2})
  +r^{2}V
  ~,
\end{eqnarray}
where we defined auxiliary scalar fields~\cite{gerosa2016numerical},
\begin{eqnarray}
  \psi
  :=
  \frac{1}{\alpha}
  \partial_{t}\varphi
  ~,~~
  \eta
  :=
  \frac{1}{X}
  \partial_{r}\varphi
  ~.
\end{eqnarray}

The equations of motion for the scalar field Eq.~(\ref{Eq:Equations-Of-Motion_scalar})
can be reformulated in a first-order system as~\cite{gerosa2016numerical,rosca2020core}
\begin{eqnarray}
  \label{Eq:EOM_Spherical_scalar_1}
  \partial_{t}\varphi
  &=&
  \alpha\psi
  ~,
  \\
  \label{Eq:EOM_Spherical_scalar_2}
  \partial_{t}\psi
  &=&
  \frac{1}{r^{2}X}
  \partial_{r}(r^{2}\alpha\eta)
  -rX\alpha\psi
  \biggr(
  \eta\psi
  +\frac{4\pi}{\alpha X}
  \tilde{T}_{tr}
  \biggr)
  +\alpha\psi^{2}
  \frac{\partial_{\varphi}F}{2F}
  \nonumber\\
  &-&2\pi\alpha
  \tilde{T}
  \partial_{\varphi}F
  -\alpha F\partial_{\varphi}V
  ~,
  \\
  \label{Eq:EOM_Spherical_scalar_{3}}
  \partial_{t}\eta
  &=&
  \frac{1}{X}
  \partial_{r}(\alpha\psi)
  -rX\alpha\eta
  \biggr(
  \eta\psi
  +\frac{4\pi}{\alpha X}
  \tilde{T}_{tr}
  \biggr)
  +\alpha\psi\eta
  \frac{\partial_{\varphi}F}{2F}
  ~.\nonumber\\
\end{eqnarray}

For the conformal factor $F$ and the potential $V$,
we follow~\cite{rosca2019inverse,huang2021scalar}.
The explicit forms are
\begin{eqnarray}
  \label{Eq:conformal_factor}
  F(\varphi)
  &=&
  \mathrm{exp}
  (
  -2\alpha_{0}\varphi
  -\beta_{0}\varphi^{2}
  )
  ~,
  \\
  \label{Eq:qubic_potential}
  V_{3}(\varphi)
  &=&
  \frac{1}{2}
  \omega_{*}^{2}
  (\varphi^{2}
  +\lambda_{3}|\varphi|^{3})
  ~,
  \\
  \label{Eq:quartic_potential}
  V_{4}(\varphi)
  &=&
  \frac{1}{2}
  \omega_{*}^{2}
  \biggr(
  \varphi^{2}
  +\frac{\lambda_{4}}{2}
  \varphi^{4}
  \biggr)
  ~.
\end{eqnarray}
Here, $\alpha_{0}$ and $\beta_{0}$ are dimensionless coupling parameters,
which determine the deviation from GR at the first post-Newtonian
order~\cite{damour1992tensor,damour1996tensor}.
The characteristic frequency associated with the scalar-field mass is given by
\begin{equation}
  \omega_{*}:=\frac{\mu}{\hbar}
  ~,
\end{equation}
where $\hbar$ is the Planck constant.
$\lambda_{3}$ and $\lambda_{4}$ are dimensionless parameters
that determine the strength of the cubic and quartic self-interactions, respectively.
Note that our definition of $\lambda_{4}$ is the same as that in Ref.~\cite{rosca2019inverse} but different from that in Ref.~\cite{cheong2019numerical},
in which the dimensional quantity $\lambda_{4}^{\prime}:=\omega_{*}^{2}\lambda_{4}/4$ is used instead of $\lambda_{4}$.

\subsection{Matter field}

The equations of motion for the matter fields in the Einstein frame
are obtained from the Bianchi identity as
\begin{eqnarray}
  \label{Eq:Equations-Of-Motion_Matter}
  \tilde{\nabla}_{\lambda}
  \tilde{T}^{\mu\lambda}
  =
  -\frac{1}{2}
  \tilde{T}
  (\partial_{\varphi}\ln F)
  \tilde{g}^{\mu\lambda}
  \partial_{\lambda}\varphi
  ~.
\end{eqnarray}
For the matter field,
we adopt a perfect fluid in the Jordan frame,
for which the energy-momentum tensor is given by
\begin{eqnarray}
  \label{Eq:Fluid_Energy-Momentum}
  T^{\mu\nu}
  =
  \rho hu^{\mu}u^{\nu}
  +Pg^{\mu\nu}
  ~,
\end{eqnarray}
with the density $\rho$,
the pressure $P$,
the specific enthalpy $h$,
and the four-velocity $u^{\mu}$.
In this case, we also solve the mass conservation law in the Jordan frame,
\begin{eqnarray}
  \label{Eq:mass-conservation}
  \nabla_{\mu}(\rho u^{\mu})=0
  ~.
\end{eqnarray}
Here, $\nabla_{\mu}$ is the covariant derivative associated with $g_{\mu\nu}$.
In spherical symmetry,
the four-velocity is given as follows:
\begin{eqnarray}
  \label{Eq:4-velocity}
  u^{\mu}
  =
  W[
  \alpha^{-1},
  v/X,
  0,
  0
  ]
  ~,
\end{eqnarray}
where $v$ is the radial velocity and $W=(1-v^{2})^{-1/2}$ is the Lorentz factor.

The equations of motion for the matter field Eq.~(\ref{Eq:Equations-Of-Motion_Matter}) are basically those of the standard general relativistic hydrodynamics in spherical symmetry, which can be transformed into a
flux conservative form by introducing the conserved variables~\cite{gerosa2016numerical},
\begin{eqnarray}
  \label{Eq:Conservative}
  D
  &:=&
  \rho WXF^{-3/2}
  ~,
  \\
  S
  &:=&
  \rho hW^{2}vF^{-2}
  ~,
  \\
  \tau
  &:=&
  (\rho hW^{2}-P)F^{-2}-D
  ~,
\end{eqnarray}
as
\begin{eqnarray}
  \label{Eq:Equations-Of-Motion_Hydro}
  &&\partial_{t}
  \left[\begin{array}{c}
  D
  \\
  S
  \\
  \tau
  \end{array}\right]
  +\frac{1}{r^{2}}
  \partial_{r}
  \biggr\{
  \frac{\alpha r^{2}}{X}
  \left[\begin{array}{c}
  Dv
  \\
  Sv+P/F^{2}
  \\
  S-Dv
  \end{array}\right]
  \biggr\}
  =
  \left[\begin{array}{c}
  s_{D}
  \\
  s_{S}
  \\
  s_{\tau}
  \end{array}\right]
  ~,
  \nonumber\\
\end{eqnarray}
where source terms are
\begin{eqnarray}
  \label{Eq:Source}
  &&
  s_{D}
  :=
  -\alpha D
  (\psi+v\eta)
  \frac{\partial_{\varphi}F}{2F}
  ~,
  \\&&
  s_{S}
  :=
  \alpha XF
  (Sv-\tau-D)
  \biggr(
  8\pi r\frac{P}{F^{2}}
  +\frac{m}{r^{2}}
  -\frac{\partial_{\varphi}F}{2F^{2}X}\eta
  -rV
  \biggr)
  \nonumber\\&&~~~
  +\frac{\alpha X}{F}
  \frac{m}{r^{2}}P
  +\frac{2\alpha P}{rXF^{2}}
  -2r\alpha XS\eta\psi
  -\frac{3}{2}
  \alpha\eta
  \frac{P}{F^{2}}
  \frac{\partial_{\varphi}F}{F}
  \nonumber\\&&~~~
  -\frac{1}{2}
  r\alpha X
  (\eta^{2}+\psi^{2})
  \biggr(
  \tau+D+\frac{P}{F^{2}}
  \biggr)
  (1+v^{2})
  \nonumber\\&&~~~
  -\alpha X\frac{P}{F}rV
  ~,
  \\&&
  s_{\tau}
  :=
  -r\alpha X
  \biggr(
  \tau+D+\frac{P}{F^{2}}
  \biggr)
  \biggr[
  (1+v^{2})
  \eta\psi
  +v(\eta^{2}+\psi^{2})
  \biggr]
  \nonumber\\&&~~~~
  +\alpha
  \frac{\partial_{\varphi}F}{2F}
  \biggr[
  Dv\eta
  +\biggr(
  Sv-\tau
  +3\frac{P}{F^{2}}
  \biggr)
  \psi
  \biggr]
  ~.
\end{eqnarray}

To close the system,
we adopt a hybrid equation of state (EOS)~\cite{janka1993does,dimmelmeier2002relativistic},
following the previous studies~\cite{rosca2019inverse,sperhake2017long,rosca2020core,cheong2019numerical,huang2021scalar,gerosa2016numerical}.
This EOS consists of a cold part $P_\mathrm{cold}$ and a thermal part $P_\mathrm{th}$ as
\begin{eqnarray}
  \label{Eq:Hybrid_EOS}
  P
  =
  P_\mathrm{cold}
  +P_\mathrm{th}
  ~.
\end{eqnarray}
Here, the cold part is written as the piecewise polytropic EOS,
\begin{eqnarray}
  \label{Eq:Cold_Part}
  P_\mathrm{cold}
  =
  \left\{\begin{array}{cc}
  \displaystyle K_{1}\rho^{\Gamma_{1}}
  &
  \rho\le\rho_\mathrm{nuc}
  \\
  \displaystyle K_{2}\rho^{\Gamma_{2}}
  &
  \rho>\rho_\mathrm{nuc}
  \end{array}\right.
  ~,
\end{eqnarray}
and the thermal part is the $\Gamma$-law type EOS given by
\begin{eqnarray}
  \label{Eq:Thermal_Part}
  P_\mathrm{th}
  =
  (\Gamma_\mathrm{th}-1)
  \rho(\epsilon-\epsilon_\mathrm{cold})
  ~,
\end{eqnarray}
where $\rho_\mathrm{nuc}=2\times10^{14}~\mathrm{g/cm^{3}}$ is the nuclear density,
$\epsilon$ is the specific internal energy,
and $\epsilon_\mathrm{cold}$ is its cold part,
\begin{eqnarray}
  \epsilon_\mathrm{cold}
  =
  \left\{\begin{array}{lc}
  \displaystyle
  \frac{K_{1}\rho^{\Gamma_{1}-1}}{\Gamma_{1}-1}
  &
  \rho\le\rho_\mathrm{nuc}
  \\
  \displaystyle
  \frac{K_{2}\rho^{\Gamma_{2}-1}}{\Gamma_{2}-1}
  +\frac{K_{1}\rho_\mathrm{nuc}^{\Gamma_{1}-1}}{\Gamma_{1}-1}
  -\frac{K_{2}\rho_\mathrm{nuc}^{\Gamma_{2}-1}}{\Gamma_{2}-1}
  &
  \rho>\rho_\mathrm{nuc}
  \end{array}\right.
  .
  \nonumber\\
\end{eqnarray}

We set $K_{1}=4.9345\times10^{14}~\mathrm{cgs}$~\cite{shapiro2008black},
and $K_{2}$ is obtained from the continuity of the pressure at $\rho=\rho_\mathrm{nuc}$.
Parameters of the hybrid EOS are $(\Gamma_{1},\Gamma_{2},\Gamma_\mathrm{th})$.
Because spectra of scalar GWs depend weakly on the EOS~\cite{cheong2019numerical} so that
we fix $(\Gamma_{1},\Gamma_{2},\Gamma_\mathrm{th})=(1.3,2.5,1.35)$ in all simulations
following previous studies~\cite{rosca2019inverse,cheong2019numerical,huang2021scalar}.

\section{
SCALAR MODES OF GRAVITATIONAL WAVES: PROPAGATION AND OBSERVATION
}
\label{sec:SGW}
\subsection{Propagation of scalar GWs}

In the wave zone,
the propagation equation of scalar GWs in the quartic self-interaction
is well approximated by~\cite{cheong2019numerical,rosca2019inverse}
\begin{eqnarray}
  \label{Eq:wave-equation-interaction}
  \partial_{t}^{2}\sigma
  -\partial_{r}^{2}\sigma
  +\omega_{*}^{2}\sigma
  +\lambda_{4}
  \omega_{*}^{2}
  \frac{\sigma^{3}}{r^{2}}
  =
  0
  ~,
\end{eqnarray}
where $\sigma:= r\varphi$ is the rescaled scalar field.
Note that the interaction term decays as $r^{-2}$ (in the case of the cubic self-interaction, $r^{-1}$~\cite{huang2021scalar}).
Therefore, if we extract the scalar field at $r_\mathrm{ex}$, which is sufficiently far away from the source of GWs,
the mass term dominates, and subsequent evolution obeys a simple wave equation in the flat spacetime,
\begin{eqnarray}
  \label{Eq:wave-equation}
  \partial_{t}^{2}\sigma
  -\partial_{r}^{2}\sigma
  +\omega_{*}^{2}\sigma
  =
  0
  ~.
\end{eqnarray}
Following~\cite{cheong2019numerical},
we set this extraction radius for the quartic self-interaction as
$r_\mathrm{ex}\approx2.53\lambda_{C}$,
where $\lambda_{C}\approx1.97\times10^{4}(\mu/10^{-14}~\mathrm{eV})^{-1}~\mathrm{km}$ is
the reduced Compton wavelength of the scalar field.
Note that with this value of $r_\mathrm{ex}$, it has been shown~\cite{rosca2019inverse} that
the nonlinear term in the propagation equation (\ref{Eq:wave-equation-interaction}) is small enough so that we may use Eq.~(\ref{Eq:wave-equation}) for the subsequent propagation.

For the cubic self-interaction, we also adopt the same extraction radius $r_\mathrm{ex}\approx2.53\lambda_{C}$ following
Ref.~\cite{huang2021scalar}.
We confirmed that observed GW spectra do not change drastically
when we adopt the extraction radius of $10r_\mathrm{ex}$.

Equation~(\ref{Eq:wave-equation}) gives the dispersion relation and the group velocity,
\begin{eqnarray}
  \omega^{2}
  &=&
  k^{2}+\omega_{*}^{2}
  ~,\\
  v_{g}
  &=&
  \frac{d\omega}{dk}
  =
  \sqrt{
  1-(\omega_{*}/\omega)^{2}}
  ~.
\end{eqnarray}
It is found that low frequency ($\omega<\omega_{*}$) modes decay exponentially and higher frequency modes
propagate faster (so-called inverse chirp~\cite{sperhake2017long,rosca2020core}).
For an event at a typical astrophysical-scale distance, the time lag between different frequency modes will
stretch out, and consequently, scalar GWs will be detected as quasimonochromatic signals, lasting for many years~\cite{sperhake2017long,rosca2020core}.
Note that these characteristic features of scalar GWs
(inverse chirp, quasimonochromatic, and long-lived)
are caused by the scalar-field mass~\cite{sperhake2017long,rosca2020core}.

The dispersive nature of scalar GWs causes difficulties in
numerical calculations~\cite{sperhake2017long,rosca2020core}.
Scalar GWs at the observation radius $D_\mathrm{obs}$ are significantly different from those at
the extraction radius $r_\mathrm{ex}$ adopted in the numerical simulations.
This means that, unlike usual tensor modes of GWs, we can not simply regard scalar GWs extracted at $r_\mathrm{ex}$
as the observed signals.
To overcome this, we employ the stationary phase approximation to propagate $\sigma(t,r_\mathrm{ex})$
toward $D_\mathrm{obs}$~\cite{sperhake2017long,rosca2020core}.
Using this procedure, the scalar GW at $D_\mathrm{obs}$ is given as~\cite{sperhake2017long,rosca2020core}
\begin{eqnarray}
  \sigma(t,D_\mathrm{obs})
  &=&
  A(t,D_\mathrm{obs})
  \cos[\phi(t,D_\mathrm{obs})]
  ~,
  \\
  \label{Eq:Observed_Amplitude}
  A(t,D_\mathrm{obs})
  &=&
  \sqrt{
  \frac{2(\Omega^{2}(t)-\omega_{*}^{2})^{3/2}}
  {\pi\omega_{*}^{2}(D_\mathrm{obs}-r_\mathrm{ex})}
  }
  |\tilde{\sigma}(\Omega,r_\mathrm{ex})|
  ~,
  \\
  \phi(t,D_\mathrm{obs})
  &=&
  -\Omega(t)t
  +\sqrt{\Omega^{2}(t)-\omega_{*}^{2}}
  (D_\mathrm{obs}-r_\mathrm{ex})
  \nonumber\\&&
  -\frac{\pi}{4}
  +\mathrm{arg}
  \{
  \tilde{\sigma}(\Omega,r_\mathrm{ex})
  \}
  ~,
\end{eqnarray}
where $\tilde{\sigma}$ is the Fourier transform of $\sigma$.
$\Omega(t)$ is the characteristic frequency arising from the stationary phase approximation
and is given by~\cite{sperhake2017long,rosca2020core}
\begin{eqnarray}
  \label{Eq:characteristic frequency}
  \Omega(t)
  =
  \frac{\omega_{*}}
  {\sqrt{1-[(D_\mathrm{obs}-r_\mathrm{ex})/t]^{2}}}
  ~
  (t>D_\mathrm{obs}-r_\mathrm{ex})
  ~.~
\end{eqnarray}
In this work, we set $D_\mathrm{obs}=10~\mathrm{kpc}$ following~\cite{rosca2020core,huang2021scalar}.
Note that $\Omega(t)$ shows the inverse chirp structure:
it reaches $\omega_{*}$ as $t\to\infty$, and low frequency modes $(\omega<\omega_{*})$ never reach $D_\mathrm{obs}$.

\subsection{Scalar GW observations}

The strain amplitude of the scalar GW consists of
a breathing mode $h^{b}=2\alpha_{0}\varphi$
and a longitudinal mode $h^{l}=(\omega_{*}/\omega)^{2}h^{b}$~\cite{sperhake2017long,rosca2020core,huang2021scalar}.
Antenna pattern functions for scalar modes are
$F_{s}(\theta,\phi):= F_{b}=-F_{l}=-\sin^{2}\theta\cos(2\phi)/2$,
which depends on the sky location $(\theta,\phi)$ of the source of GWs~\cite{will2014confrontation}.
As a result, the scalar GW signal is
\begin{eqnarray}
  \label{Eq:strain_amplitude}
  h(t)
  =
  F_{s}(h^{b}-h^{l})
  =
  2\alpha_{0}
  F_{s}
  \biggr[
  1
  -\biggr(
  \frac{\omega_{*}}{\omega}
  \biggr)^{2}
  \biggr]
  \varphi
  ~.
\end{eqnarray}
In this work,
we used the sky-location-averaged root mean square value~\cite{rosca2020core} for simplicity,
\begin{eqnarray}
  \bar{F}_{s}
  :=
  \sqrt{
  \int d\Omega~
  F^{2}_{s}(\theta,\phi)
  }
  =
  \sqrt{
  \frac{4\pi}{15}
  }
  ~.
\end{eqnarray}

The signal to noise ratio $\rho$ is defined as
\begin{eqnarray}
  \rho^{2}
  =
  4\int^{\infty}_{0}df~
  \frac{|\tilde{h}(f)|^{2}}{S_{n}(f)}
  ~,
\end{eqnarray}
where $\tilde{h}(f)$ is the Fourier transform of $h(t)$ and
$S_{n}(f)$ is the one-sided noise power spectral density.
For a (quasi)monochromatic signal, the signal to noise ratio is approximated as
\begin{eqnarray}
  \rho
  \approx
  \sqrt{\frac{S_{o}(\Omega/2\pi)}{S_{n}(\Omega/2\pi)}}
  ~,
\end{eqnarray}
where $\sqrt{S_{o}}$ is the power spectral density of the scalar GW signal
and is given by~\cite{rosca2020core,huang2021scalar}
\begin{eqnarray}
  \label{Eq:observed_spectra}
  \sqrt{S_{o}\biggr(\frac{\Omega}{2\pi}\biggr)}
  :=
  \alpha_{0}
  \sqrt{T}
  \bar{F}_{s}
  \frac{A(t,D_\mathrm{obs})}{D_\mathrm{obs}}
  \biggr[
  1
  -\biggr(
  \frac{\omega_{*}}{\Omega}
  \biggr)^{2}
  \biggr]
  ~.
\end{eqnarray}
Here, $T$ is the duration of an observation
and we set $T=2$ months following~\cite{rosca2020core,huang2021scalar}.

\section{
Numerical Setup
\label{sec:Setup}
}
\begin{figure}[b!]
  \includegraphics
  [clip,keepaspectratio,width=7cm]
  {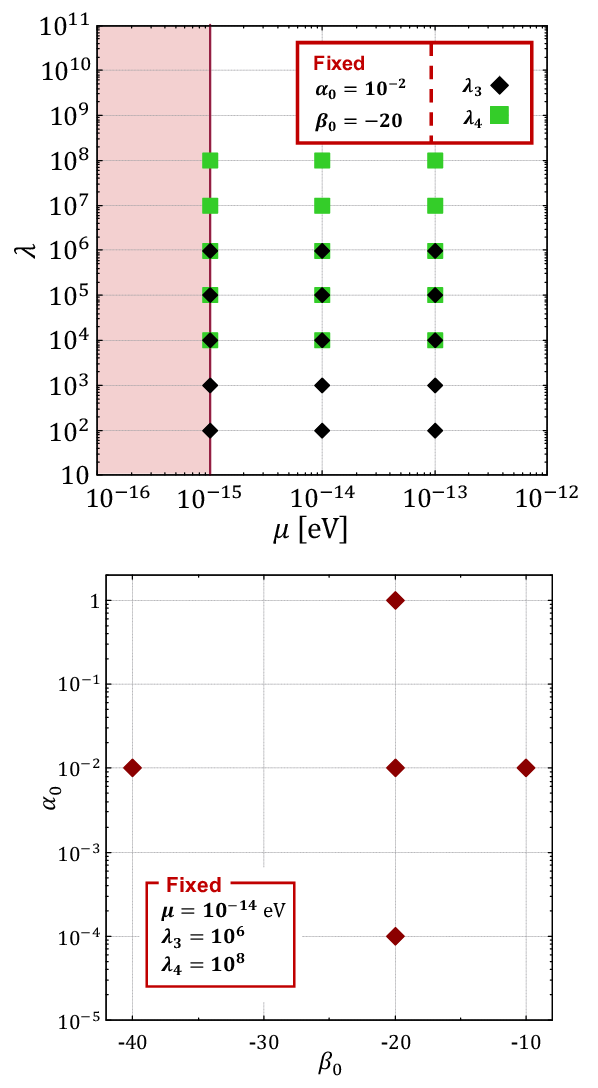}
  \caption{
  Top panel:
  the range of $\mu$-$\lambda_{i}$ space
  with fixed $(\alpha_{0},\beta_{0})=(10^{-2},-20)$.
  Black diamonds and green boxes show values of $\lambda_{3}$ and $\lambda_{4}$, respectively.
  The red-filed area shows the parameter region,
  where observational constraints from binary pulsars should be considered.
  Bottom panel:
  the range of $\beta_{0}$-$\alpha_{0}$ space
  with fixed $(\mu,\lambda_{3},\lambda_{4})=(10^{-14}~\mathrm{eV},10^{6},10^{8})$.
  }
  \label{Fig:Parameter}
\end{figure}

\begin{figure*}[!]
  \includegraphics
  [clip,width=16cm]
  {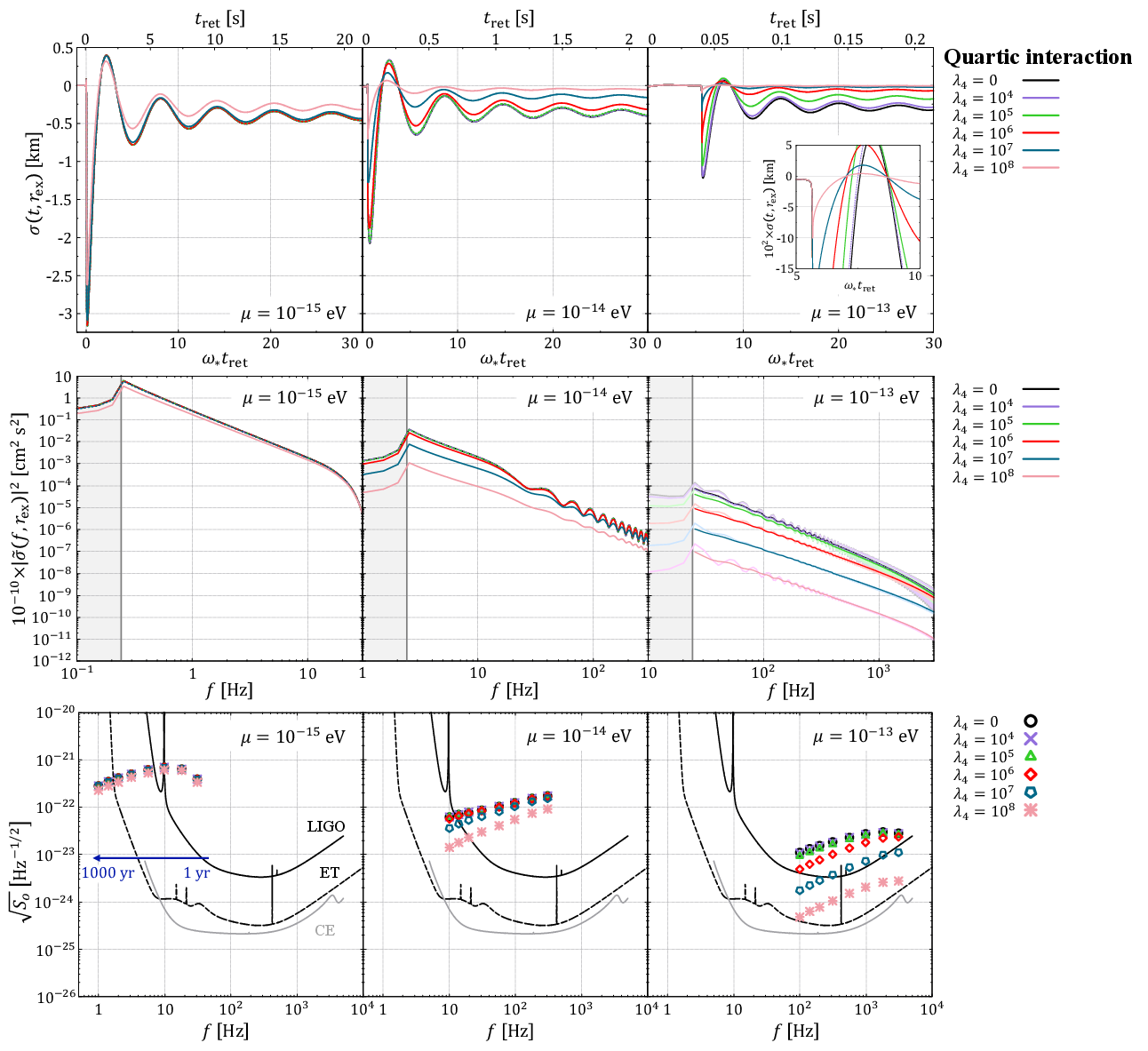}
  \caption{
  Time evolution of the scalar field $\sigma(t,r_\mathrm{ex})$ at
  $r_\mathrm{ex}\approx2.53\lambda_{C}$ (top panels),
  Fourier spectra $\tilde{\sigma}(f,r_\mathrm{ex})$ of $\sigma(t,r_\mathrm{ex})$ (middle panels),
  and observed spectra $\sqrt{S_{o}(f)}$ at $D_\mathrm{obs}=10~\mathrm{kpc}$ (bottom panels)
  for the quartic self-interaction.
  The scalar-field mass is $\mu=10^{-15},10^{-14}$, and $10^{-13}~\mathrm{eV}$ from left to right panels,
  and parameters of the conformal factor are $(\alpha_{0},\beta_{0})=(10^{-2},-20)$.
  Top panels:
  the lower horizontal axes are the retarded time normalized by $\omega_{*}$,
  and those ranges are the same for each scalar-field mass $0\le\omega_{*}t_\mathrm{ret}\le30$.
  The upper horizontal axes are the retarded time.
  Middle panels:
  for models with $\mu=10^{-13}~\mathrm{eV}$,
  averaged spectra (thin curves) are shown together.
  The gray-filed area $(f<\omega_{*}/2\pi=2.42(\mu/10^{-14}~\mathrm{eV})~\mathrm{Hz})$ shows nonpropagating modes.
  Bottom panels:
  sensitivity curves of GW detectors are also plotted:
  Advanced LIGO (black line),
  Einstein Telescope (black dashed line),
  and Cosmic Explorer (gray line).
  Symbols represent different observer's retarded times:
  $t_\mathrm{obs,ret}=1,3,10,30,100,250,500$, and $1000~\mathrm{yr}$
  from right to left on each plot following~\cite{rosca2020core,huang2021scalar}.
  }
  \label{Fig:quartic_waveforms_spectra}
\end{figure*}

\begin{figure*}[!]
  \includegraphics
  [clip,width=16cm]
  {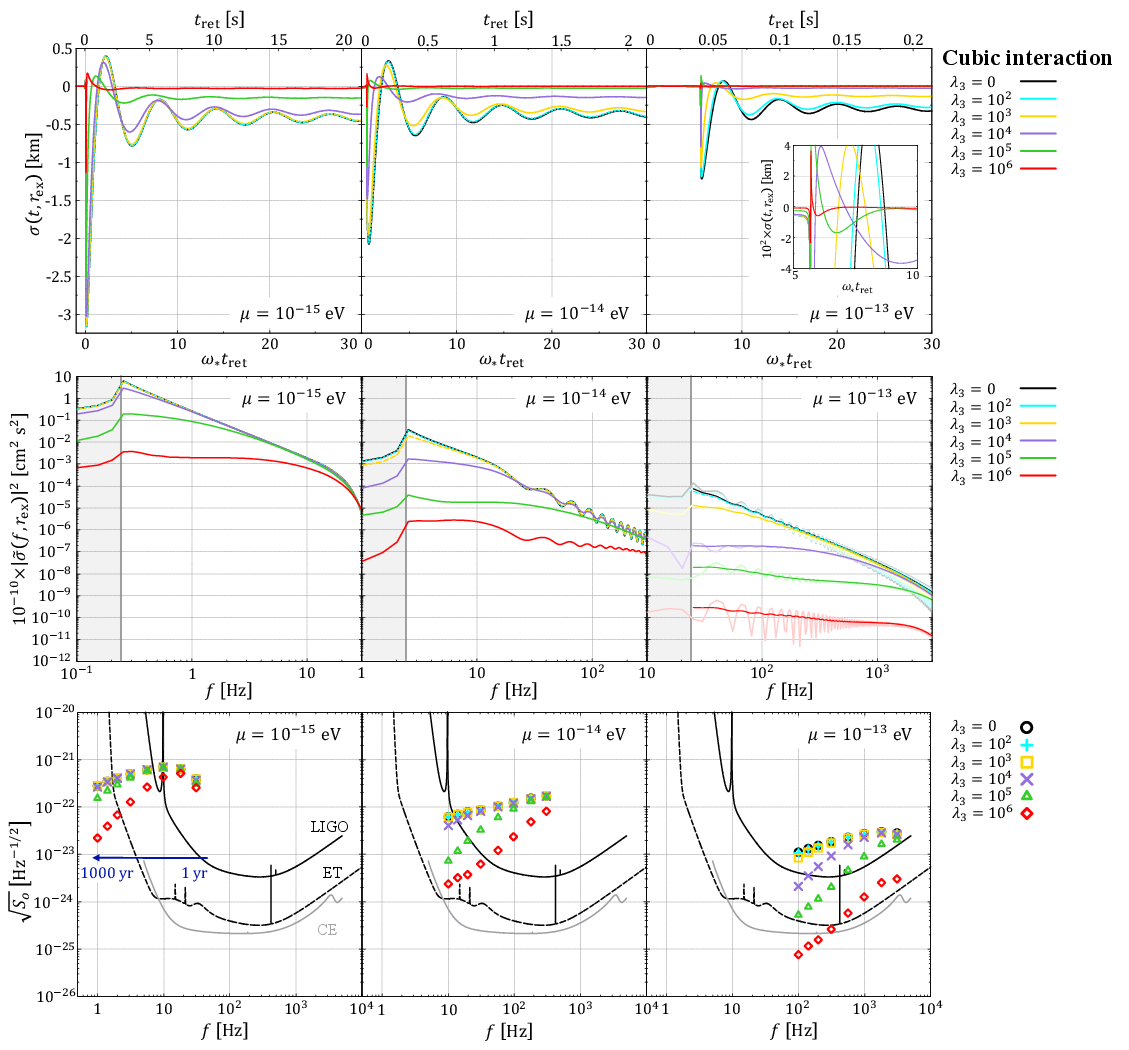}
  \caption{
  Similar to Fig.~\ref{Fig:quartic_waveforms_spectra},
  but for the cubic self-interaction.
  }
  \label{Fig:cubic_waveforms_spectra}
\end{figure*}

\begin{figure*}[!]
  \includegraphics
  [clip,keepaspectratio,width=15cm]
  {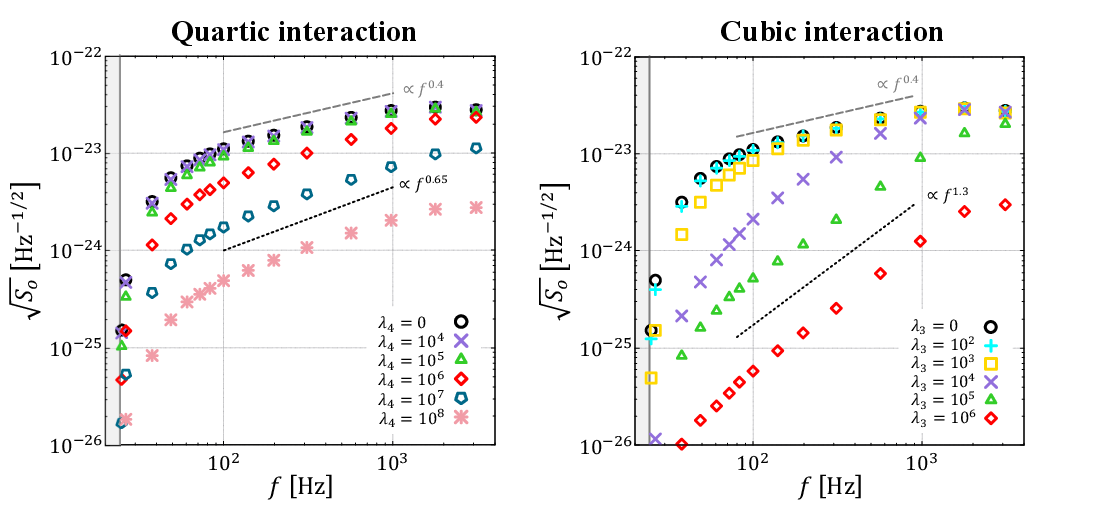}
  \caption{
  Observed spectra for the quartic self-interaction (left panel) and the cubic self-interaction (right panel).
  The scalar-field mass is $\mu=10^{-13}~\mathrm{eV}$.
  The gray-filed area $(f<\omega_{*}/2\pi=24.2~\mathrm{Hz})$ shows the no scalar GW signals area.
  Since low frequency modes $(\omega<\omega_{*})$ never reach $D_\mathrm{obs}$,
  observed spectra damp exponentially near the cutoff frequency $\omega_{*}$.
  }
  \label{Fig:slope}
\end{figure*}

In the present study, we perform two sets of simulations.
In the first set, we fix the parameters of the conformal factor [see Eq.~(\ref{Eq:conformal_factor})]
to be $(\alpha_{0},\beta_{0})=(10^{-2}, -20)$
following previous studies~\cite{cheong2019numerical,huang2021scalar},
while we varied the scalar-field mass $\mu$ and the self-interaction parameters $\lambda_{i}$ of the potential
[see Eqs.~(\ref{Eq:qubic_potential}) and (\ref{Eq:quartic_potential})]
in the ranges shown in the upper panel of Fig.~\ref{Fig:Parameter}.

Taking into account the result that the amplitude of the scalar field is small as $|\varphi|<1$
(see Fig.~\ref{Fig:dynamics_mass}),
we set the range of $\lambda_{4}$ to be larger than that of $\lambda_{3}$.
In the first type of simulations, a strong
scalarization occurs in the neutron star formed after the collapse for most of the parameters $(\lambda_{i},\mu)$
as found in Refs.~\cite{cheong2019numerical,huang2021scalar} (see also Appendix \ref{Sec_Collapse-dynamics}).

For the second set, we fix the scalar-field mass and
the self-interaction parameters as $\mu=10^{-14}~\mathrm{eV}$, $\lambda_{3} = 10^{6}$, and $\lambda_{4} = 10^{8}$,
respectively, while the parameters of the conformal factor $(\alpha_{0},\beta_{0})$ are varied
in the ranges shown in the lower panel of Fig.~\ref{Fig:Parameter}
to see the dependence of scalar GWs on them.
We set these values of $(\alpha_{0},\beta_{0})$ in the range where a strong scalarization can occur in the neutron star formed after the collapse.
The adopted values of $(\alpha_{0},\beta_{0})$ satisfy the current observational constraints for the massive scalar field.
Note that the constraints are more relaxed for the massive scalar field~\cite{ramazanouglu2016spontaneous,yazadjiev2016slowly}
(see also Ref.~\cite{berti2015testing} for current constraints on the massless ST theory).

For the initial model of stellar core collapse, we adopt a nonrotating presupernova model with solar
metallicity in Ref.~\cite{woosley2002evolution}.
We use the model with the mass of $12M_{\odot}$ (this model is denoted as s12 in Ref.~\cite{rosca2020core})
and set vanishing scalar fields initially.
In the hydrodynamics simulation, we need to set the atmosphere density $\rho_\mathrm{atmo}$.
In this study, we adopt $\rho_\mathrm{atmo}=1~\mathrm{g/cm^{3}}$.

According to~\cite{rosca2020core}, we use a grid consisting of
a uniform grid with a grid spacing of $\Delta r=100~\mathrm{m}$ up to $r=40~\mathrm{km}$
and a nonuniform grid with a logarithmically spacing up to $R_\mathrm{out}$.
We set $R_\mathrm{out}=5\times10^{6}~\mathrm{km}$ for models with
$\mu=10^{-15}~\mathrm{eV}$
and $R_\mathrm{out}=5\times10^{5}~\mathrm{km}$ for the other scalar-field mass models.

Note that grid resolution at $r_\mathrm{ex}$ is important to extract scalar GWs.
In this study, the maximum frequency of scalar GWs is given by
$f_\mathrm{max}:=\Omega_\mathrm{max}/2\pi\approx3.1\times10^{2}(\mu/10^{-14}~\mathrm{eV})~\mathrm{Hz}$,
where $\Omega_\mathrm{max}$ is the characteristic frequency Eq.~(\ref{Eq:characteristic frequency}) at
the observer's retarded time $1~\mathrm{yr}$.
The minimum wavelength associated with this maximum frequency is
$\lambda_\mathrm{min}:=c/f_\mathrm{max}\approx9.7\times10^{2}(\mu/10^{-14}~\mathrm{eV})^{-1}~\mathrm{km}$.
The grid resolution which sufficiently resolves $\lambda_\mathrm{min}$ is needed,
and we set $\lambda_\mathrm{min}/\Delta r\approx100$ at $r_\mathrm{ex}$ for models with $\mu=10^{-14}$ and $10^{-13}~\mathrm{eV}$.
For models with $\mu=10^{-15}~\mathrm{eV}$,
we set $\lambda_\mathrm{min}/\Delta r\approx80$ to reduce the computation cost.
\footnote{
In this estimate,
we use the light speed instead of the exact phase velocity
$v_p=[1-(\omega_{*}/\omega)^{2}]^{-1/2}$ for simplicity.
If we use this proper phase velocity,
the minimum wavelength changes by about $3\times10^{-3}\%$.
This slight difference has no effect on the condition for $\Delta r$ at $r_\mathrm{ex}$.
}
With these values of $\Delta r$, our main results are not affected by $\Delta r$.
Note, however, that in the much higher frequency range ($f>f_\mathrm{max}$),
the amplitude of Fourier spectra damps due to a lack of grid resolution (see Appendix~\ref{Sec_Convergence}).

In reference to Refs.~\cite{gerosa2016numerical,rosca2020core},
we constructed a new numerical code
which can simulate spherical core collapse in the massive ST theory with self-interactions,
and details of the code are in those references.
Results of some test simulations will be found in Appendix~\ref{Sec_Code-tests}.

\section{
Results
\label{sec:Results}
}

\subsection{
Overall features of scalar GW signals
}

General features of collapse dynamics and evolution of the scalar field agree well with those
in the previous studies~\cite{cheong2019numerical,rosca2020core,huang2021scalar},
as briefly summarized in Appendix~\ref{Sec_Collapse-dynamics}.
As a consequence, overall features of scalar GW signals also agree well as shortly outlined below.

Figure \ref{Fig:quartic_waveforms_spectra} shows
time evolution of the scalar field at the extracted radius
$\sigma(t,r_\mathrm{ex}) = r_\mathrm{ex}\varphi(t,r_\mathrm{ex})$ (upper panels),
Fourier spectra $\tilde{\sigma}(f,r_\mathrm{ex})$ of the scalar field $\sigma(t,r_\mathrm{ex})$ (middle panels),
and observed spectra $\sqrt{S_{o}(f)}$ (lower panels)
for the quartic self-interaction with $(\alpha_{0},\beta_{0})=(10^{-2},-20)$.
As can be seen in Fig.~\ref{Fig:quartic_waveforms_spectra}, the self-interaction of the
scalar field suppresses scalar GWs, and the suppression is more prominent for the larger
self-interaction parameter $\lambda_{4}$ (see the upper and middle panels)
as found in the previous studies~\cite{cheong2019numerical,rosca2019inverse}.
We computed the observed spectra $\sqrt{S_{o}(f)}$ using the stationary phase approximation and found that the observed spectra are also suppressed more for the larger $\lambda_{4}$, with showing the inverse chirp feature.

In Fig.~\ref{Fig:quartic_waveforms_spectra}, we also show the dependence of the scalar GW signals on the scalar-field mass (compare panels in the first, second, and third columns).
The amplitude of both extracted and observed scalar GWs decrease as one increases the scalar-field mass as found in the previous study~\cite{rosca2020core}.
We also found that the observed GW spectra shift towards higher frequencies when the scalar-field mass is increased.
This is simply because the characteristic frequency $\Omega(t)$ that determines the range of observed spectra linearly shifts as one varies the scalar-field mass [see Eq.~(\ref{Eq:characteristic frequency})].

In Fig.~\ref{Fig:cubic_waveforms_spectra}, we show the results of the scalar GW signals for the cubic self-interaction. We found that the general features of the scalar GW signals are qualitatively similar to those for the quartic self-interaction. That is, the suppression of the GW amplitude is more significant for the larger self-interaction parameter $\lambda_{3}$~\cite{huang2021scalar}.
From a quantitative viewpoint, however, the suppression of the scalar GW signals depends on the type of the self-interaction in a systematic manner.
We found that the type of the self-interaction can be constrained using this systematic dependence of scalar GWs as described in the next subsection.

Note that extracted and observed scalar GW spectra for models with $\mu=10^{-15}~\mathrm{eV}$ decrease in the higher frequency range $(f>10~\mathrm{Hz})$ due to a lack of grid resolution.
However, this numerical feature does not affect our findings (see Appendix~\ref{Sec_Convergence}).

\subsection{Dependence of the suppression of scalar GWs on the self-interaction}

As noticed in Ref.~\cite{cheong2019numerical}, the spectra of the extracted scalar GW signals $\tilde{\sigma}(f, r_\mathrm{ex})$ for the models with the self-interaction are suppressed compared to that for the no self-interaction model in a frequency dependent manner (see the middle panels of Figs.~\ref{Fig:quartic_waveforms_spectra} and \ref{Fig:cubic_waveforms_spectra}).
Thanks to the comprehensive parameter study performed in this paper, we found that the
``frequency dependence" of the suppression of scalar GWs found in Ref.~\cite{cheong2019numerical} shows two characteristics as follows.
\begin{enumerate}
\item Our results suggest that the suppression of the scalar GWs in the low frequency appears
  as the bifurcation or the deviation from
  the no self-interaction case towards the lower frequency, not as overall decrease of the amplitude
  (see the middle panels of Figs.~\ref{Fig:quartic_waveforms_spectra} and~\ref{Fig:cubic_waveforms_spectra}).
  Furthermore, the frequency (hereafter noted as $f_\mathrm{bif}$) at which the bifurcation takes place depends on the parameter $\lambda_{i}$: $f_\mathrm{bif}$ is higher for the larger $\lambda_{i}$.
\item After the bifurcation, the GW spectra decline towards the lower frequency. Our results
  indicate that
  the ``frequency dependence", that is, the slope of the declined spectra depends
  systematically on the type of the self-interaction but not
  on the parameter $\lambda_{i}$ within the same self-interaction.
  This is likely to because it is the power-law index of the self-interaction that dominates the slope of the spectra as in the no self-interaction case.
\end{enumerate}

\begin{figure*}[!]
  \includegraphics
  [clip,keepaspectratio,width=14cm]
  {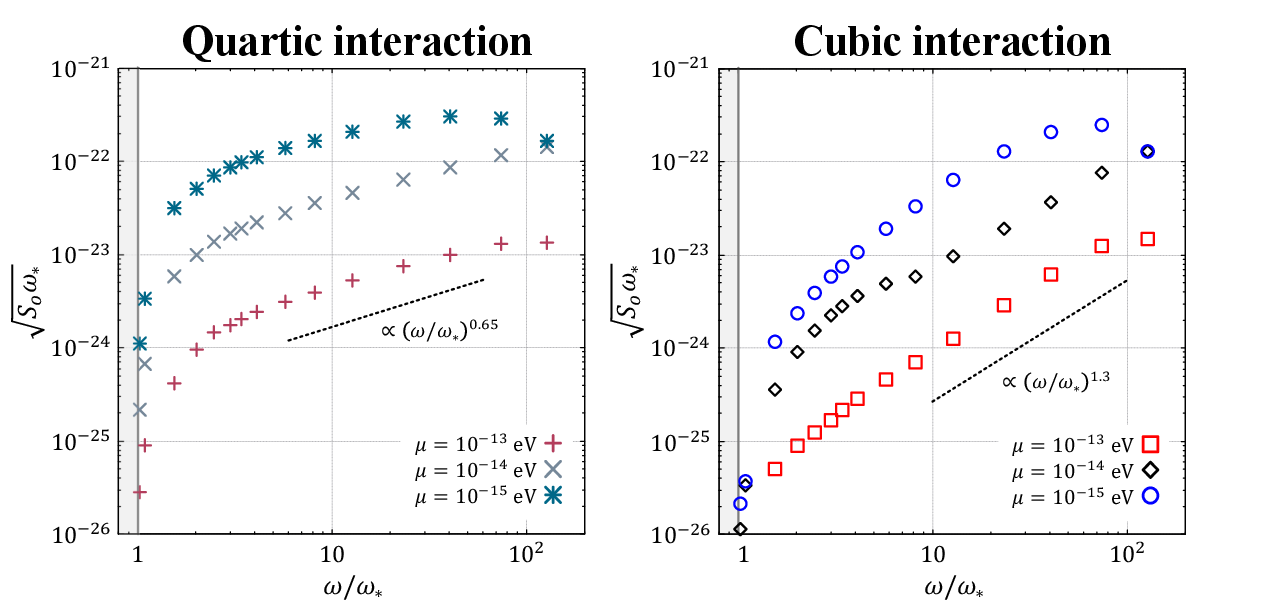}
  \caption{
  Rescaled observed spectra $\sqrt{S_{o}\omega_{*}}$ with $\lambda_{4}=10^{8}$ for the quartic self-interaction (left panel)
  and $\lambda_{3}=10^{6}$ for the cubic self-interaction (right panel).
  The horizontal axis is the rescaled frequency $\omega/\omega_{*}$.
  The gray-filed area $(\omega/\omega_*<1)$ shows the no scalar GW signals area.
  }
  \label{Fig:rescaled}
\end{figure*}

\begin{figure*}[!]
  \includegraphics
  [clip,keepaspectratio,width=15cm]
  {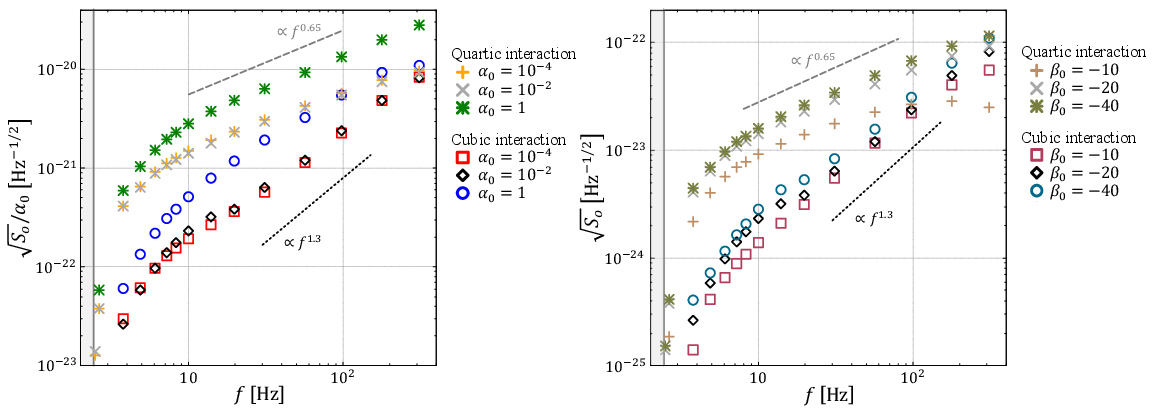}
  \caption{
  Observed spectra
  with fixed $\mu=10^{-14}~\mathrm{eV}$, $\lambda_{4}=10^{8}$ for the quartic self-interaction, and $\lambda_{3}=10^{6}$ for the cubic self-interaction, while $(\alpha_{0},\beta_{0})$ are varied.
  The gray-filed area $(f<\omega_{*}/2\pi=2.42~\mathrm{Hz})$ shows the no scalar GW signals area.
  Left panel:
  observed spectra rescaled by $\alpha_{0}$ with varied $\alpha_{0}$ and fixed $\beta_{0}=-20$.
  The rescaling of $\sqrt{S_{o}}$ by $\alpha_{0}$ eliminates the effect of the shift of the amplitude of $\sqrt{S_{o}}$
  which appears when $\alpha_{0}$ is varied [see Eq. (\ref{Eq:observed_spectra})].
  Right panel:
  observed spectra with varied $\beta_{0}$ and fixed $\alpha_{0}=10^{-2}$.
  }
  \label{Fig:conformal}
\end{figure*}

As a consequence of the above characteristic dependence of the extracted scalar GW spectra on the self-interaction,
the observed spectra also exhibit the corresponding features.
Figure \ref{Fig:slope} shows the observed spectra $\sqrt{S_{o}}$ for the models with the scalar-field mass $\mu=10^{-13}~\mathrm{eV}$.
The observed spectra also show deviation from the no self-interaction case towards the lower frequency,
and the slope of the declined observed spectra also depends systematically on the type of the self-interaction.
Note that the observed spectra show further declination near the cutoff frequency $\omega_{*}$.
We found that the declined observed spectra
in an intermediate frequency range [$O(\omega_{*}/2\pi)\lesssim f \lesssim f_\mathrm{bif}$]
can be reasonably described by a single power law as a function of the frequency irrespective of $\lambda_{i}$:
$\sqrt{S_{o}}\propto f^{0.65}$ for the quartic self-interaction and $\sqrt{S_{o}}\propto f^{1.3}$ for the cubic self-interaction.
It is remarkable that the power-law index of the GW spectra for
the cubic self-interaction is approximately twice larger than that for the quartic self-interaction.

We checked that the power-law index of the declined observed spectra is converged (see Appendix \ref{Sec_Convergence}) and does not depend on the extraction radius
if it is larger than $r_\mathrm{ex}$ adopted in this paper:
when we adopt the extraction radius $10r_\mathrm{ex}$, we confirmed that the power-law index is unchanged.
The result that the power-law index of the declined observed spectra for the cubic self-interaction is approximately twice larger than that for the quartic self-interaction does not depend on the scalar-field mass and the conformal factor as shown in the next subsection.
We also explored the impact of the progenitor and EOS on the power-law index and found that
the power-law index is insensitive to them (see Appendix \ref{Sec_EOS_Progenitor}).
Thus, our result suggests that we could constrain the type of the self-interaction by the observation of scalar GWs.

\subsection{
Influence of the mass of the scalar field and conformal factor
}
\subsubsection{
Influence of the scalar-field mass
}

In this subsection,
we investigate the influence of the scalar-field mass on the power-law index of the declined spectra.
Since the power-law index does not depend on $\lambda_{i}$,
we focus on the specific models with $\lambda_{4}=10^{8}$ for the quartic self-interaction
and $\lambda_{3}=10^{6}$ for the cubic self-interaction.

Figure~\ref{Fig:rescaled} shows rescaled observed spectra $\sqrt{S_{o}\omega_{*}}$
for the models with $\mu=10^{-15},10^{-14}$, and $10^{-13}~\mathrm{eV}$.
The horizontal axis is the rescaled frequency $\omega/\omega_{*}$.
The rescaling of $\sqrt{S_{o}}$ and $\omega$ by $\omega_{*}$ eliminates the effect of the shift of the amplitude and the frequency range of observed spectra which appears when the scalar-field mass is varied.
From Fig.~\ref{Fig:rescaled},
we found that the power-law index does not significantly change as one varies the scalar-field mass
and the behavior of the observed spectra is still described well by $\sqrt{S_{o}}\propto f^{0.65}$ for the quartic self-interaction and $\sqrt{S_{o}}\propto f^{1.3}$ for the cubic self-interaction.

\subsubsection{
Influence of the conformal factor
}

We also investigate the influence of the conformal factor on the power-law index of the observed spectra.
In this subsection,
we present results of models with $\mu=10^{-14}~\mathrm{eV}$, $\lambda_{4}=10^{8}$ for the quartic self-interaction,
and $\lambda_{3}=10^{6}$ for the cubic self-interaction.

Figure~\ref{Fig:conformal} shows observed spectra with fixed $\beta_{0}$ and varied $\alpha_{0}$ (left panel) and
fixed $\alpha_{0}$ and varied $\beta_{0}$ (right panel).
We found that the power-law index of observed spectra is also insensitive to the conformal factor
for the broad region of parameters $(\alpha_{0},\beta_{0})$.
On the other hand, the parameters of the conformal factor have impacts on observations of scalar GWs.
The amplitude of scalar GWs directly depends on $\alpha_{0}$ [$h\propto\alpha_{0}$, see Eq.~(\ref{Eq:strain_amplitude})] and consequently, the detectability of scalar GWs also depends on it.

\section{
  Summary
  \label{sec:Conclusion}
}
We performed a comprehensive study of scalar GWs in the massive ST theory
with the cubic and quartic self-interactions.
We constructed a new numerical code in reference to Refs.~\cite{gerosa2016numerical,rosca2020core},
and all simulations in this work were done by this code.

Thanks to the comprehensive and systematic parameter study,
we found that
the effect of the self-interactions on the extracted scalar GWs appears as the characteristic declination towards the lower frequency of their spectra compared with the no self-interaction case in the low frequency range.
As a consequence, the observed spectra $\sqrt{S_{o}}$ for the models with the self-interactions also show the corresponding declination.
We furthermore found that the observed spectra in the intermediate frequency range can be well described by a single power law as $\sqrt{S_{o}}\propto f^{0.65}$ for the quartic self-interaction and $\sqrt{S_{o}}\propto f^{1.3}$ for the cubic self-interaction, irrespective of $\lambda_{i}$.
Our result that the power-law index of the observed spectra for the cubic self-interaction is approximately twice larger than that for the quartic self-interaction does not depend on the mass of the scalar field and the conformal factor.
Furthermore, we confirmed that the power-law index is also insensitive to the mass of the progenitor and EOS (see Appendix \ref{Sec_EOS_Progenitor}).
All these results indicate that we may constrain the type of the self-interaction of the scalar field if we could extract the information of the power-law index of the observed spectra in future GW observations.

The amplitude of observed spectra depends on many parameters, such as $(\mu,\lambda_{i},\alpha_{0},\beta_{0})$,
and it is difficult to determine all these parameters.
However, to constrain the type of the self-interaction, we need only to extract the power-law index of observed spectra, thanks to its insensitivity to parameters other than the self-interaction type.
Therefore,
the constraint on the type of the self-interaction could be given by the one long-term observation of scalar GWs in principle.
Note, however, that when the coefficient of the self-interaction term $\lambda_{i}$ is lower, the declination of the spectra occurs at the frequency more close to (or even less than) the cutoff frequency $\omega_{*}$.
In such a case, it is difficult to constrain the type of the self-interaction because the observed spectra damp exponentially towards the lower frequency, irrespective of the type of the self-interaction.

Since $\sqrt{S_{o}}$ decreases as $D_\mathrm{obs}^{-3/2}$
[see Eqs.~(\ref{Eq:Observed_Amplitude}) and (\ref{Eq:observed_spectra})],
for most of the models with $\mu=10^{-14}~\mathrm{eV}$,
scalar GW signals at around $100~\mathrm{Hz}$ could be detectable by Cosmic Explorer up to an events at $\sim100~\mathrm{kpc}$,
where the signal to noise ratio is $\rho\sim10$.

In this paper, we employed the stationary phase approximation to propagate scalar GWs toward $D_\mathrm{obs}$.
This method is appropriate for the no self-interaction case.
For the self-interaction case, however, it is not very clear that the stationary phase approximation gives a correct result or not.
To check this, we should solve the propagation equation (\ref{Eq:wave-equation-interaction}) to properly propagate scalar GWs.
Furthermore, we did not consider rotation and magnetic fields, which may play important roles in this paper.
In addition, we focus on the stellar core collapse to a neutron star in this paper.
The case of collapse to a black hole remains to be explored.
These issues are left for future work.

\begin{figure}[b!]
  \includegraphics
  [clip,keepaspectratio,width=7cm]
  {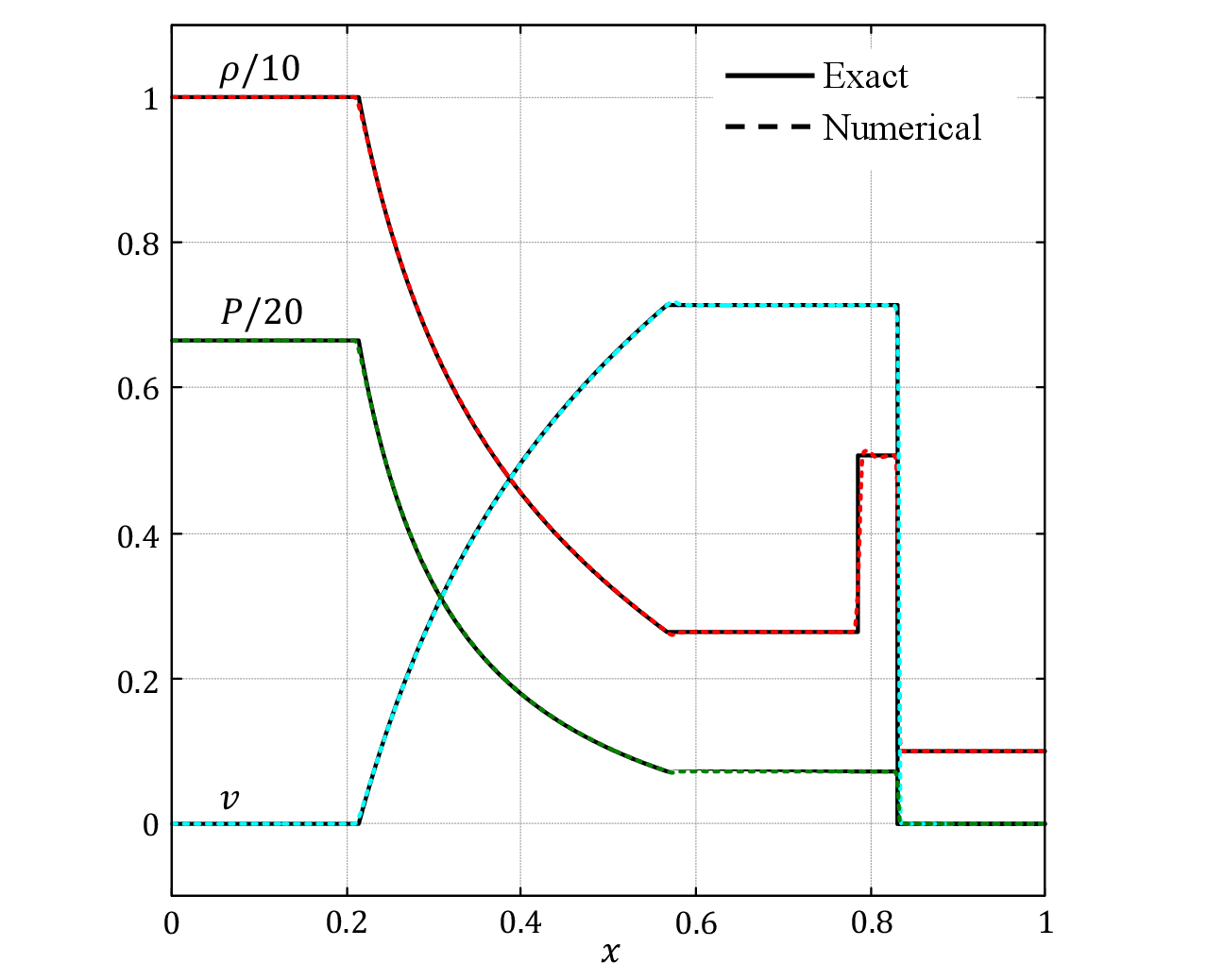}
  \caption{
  The density (red), pressure (green), and velocity (cyan) at $t=0.4$.
  Solid and dashed lines represent exact and numerical solutions, respectively.
  The density and pressure are normalized by 10 and 20, respectively.
  }
  \label{Fig:STP}
\end{figure}
\begin{figure*}[!]
  \includegraphics
  [clip,keepaspectratio,width=14cm]
  {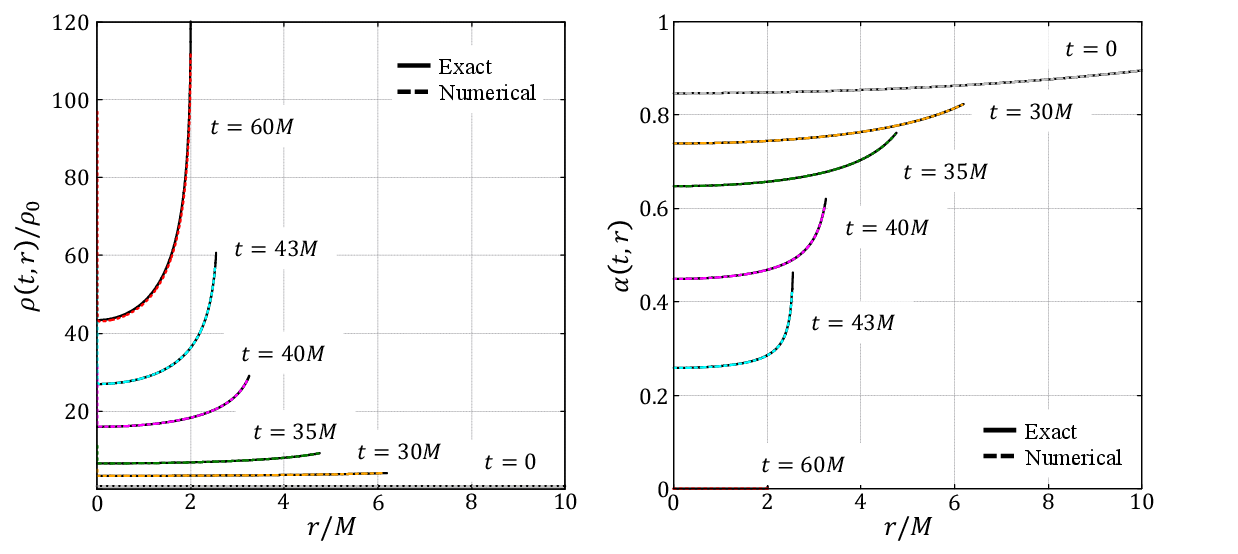}
  \caption{
  Snap shots of the density (left panel)
  and $\alpha$ (right panel).
  Solid and dashed lines represent exact and numerical solutions, respectively.
  The density $\rho(t,r)$ is normalized by the initial value $\rho_{0}:=\rho(t=0,r)$.
  }
  \label{Fig:OSC}
\end{figure*}

\begin{acknowledgments}

This work was supported by JSPS KAKENHI 20H00158 and Grant-in-Aid for Transformative Research Areas 23H04900.

\end{acknowledgments}

\appendix

\section{CODE TESTS}\label{Sec_Code-tests}
\begin{table}[b!]
  \caption{The initial condition for the shock tube problem~\cite{marti2003numerical}.}
  \label{Table:STP_initial}
  \centering
  \begin{tabular}{c|c|c}
    \hline
    Hydrodynamics variables&$x\le0.5$ & $x>0.5$
    \\
    \hline
    Density $\rho$ & 10 & 1
    \\
    Velocity $v$ & 0 & 0
    \\
    Pressure $P$ & 13.33 & $10^{-10}$
    \\
    \hline
  \end{tabular}
\end{table}

In this appendix,
we show results of two code tests:
(i) the shock tube problem and
(ii) the Oppenheimer-Snyder collapse.
Since these are the special and general relativistic tests,
we set vanishing scalar fields initially
and parameters of the conformal factor to be $\alpha_{0}=\beta_{0}=0$ in both tests.

\begin{figure*}[!]
  \begin{minipage}[t]{0.45\hsize}
    \centering
    \includegraphics
    [clip,keepaspectratio,width=8cm]
    {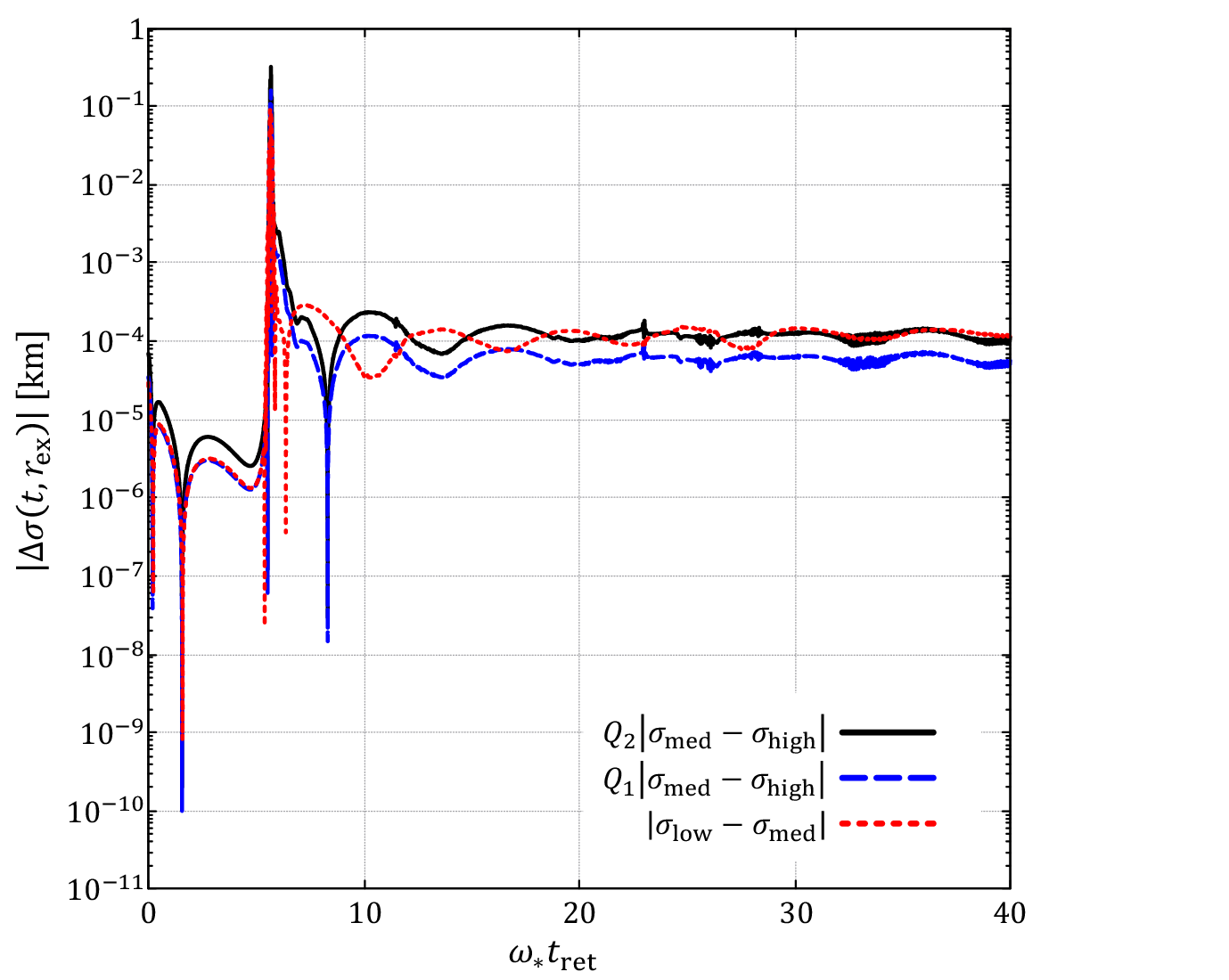}
    \caption{
    Differences between extracted scalar waveforms with different grid resolutions.
    $\sigma_{A}$ is the extracted waveform with $\Delta r_{A}$
    for $A=\mathrm{low},~\mathrm{med},~\mathrm{and~high}$.
    $Q_1$ and $Q_2$ are the first-order and second-order convergence factors, respectively.
    }
    \label{Fig:convergence}
  \end{minipage}
  \hspace{0.05\hsize}
  \begin{minipage}[t]{0.45\hsize}
    \centering
    \includegraphics
    [clip,keepaspectratio,width=8cm]
    {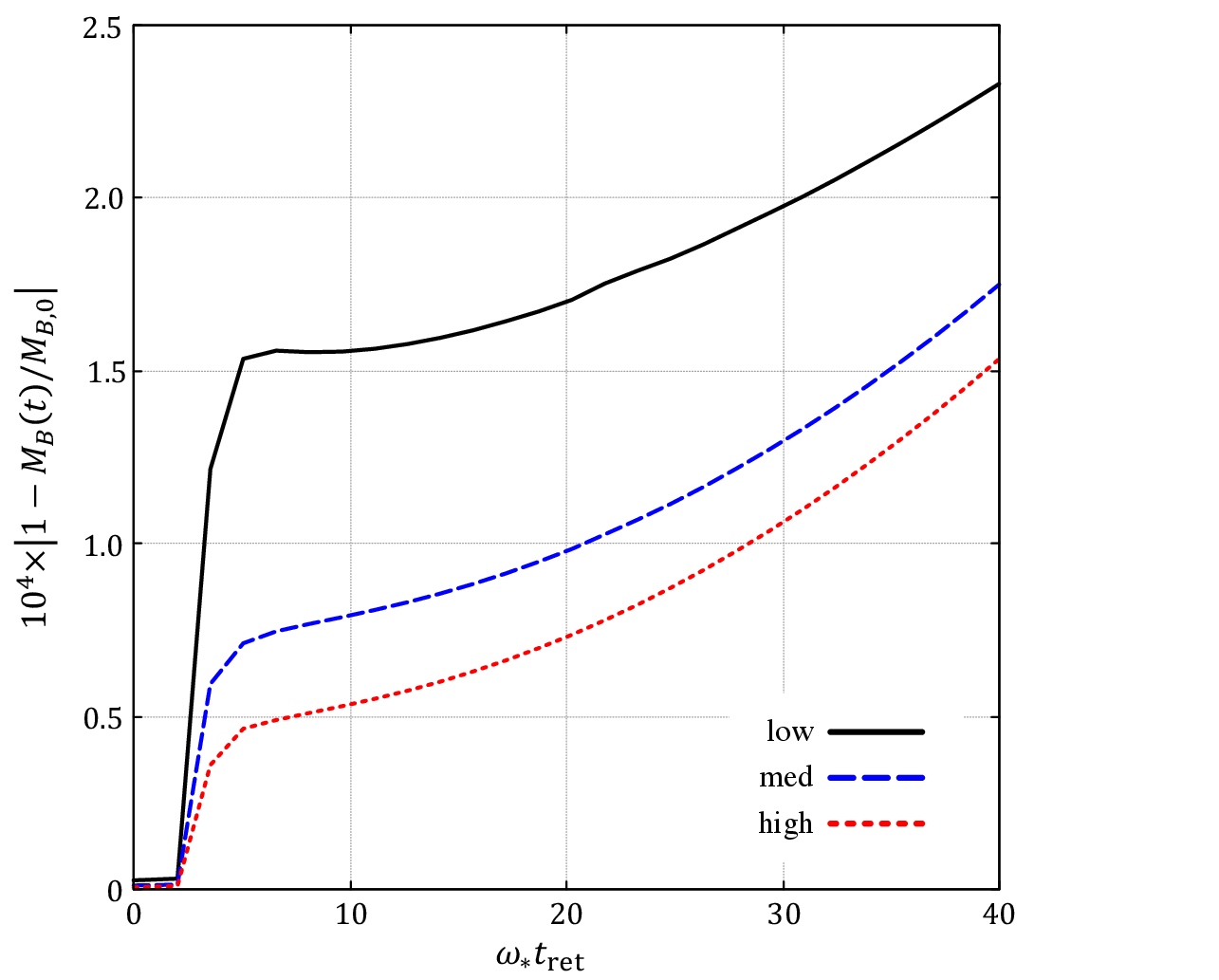}
    \caption{
    The conservation of the total baryon mass in the computation domain.
    $M_{B,0}$ is the initial total baryon mass.
    }
    \label{Fig:mass}
  \end{minipage}
\end{figure*}

\begin{figure*}[t!]
  \begin{minipage}[t]{0.45\hsize}
    \centering
    \includegraphics
    [clip,keepaspectratio,width=8cm]
    {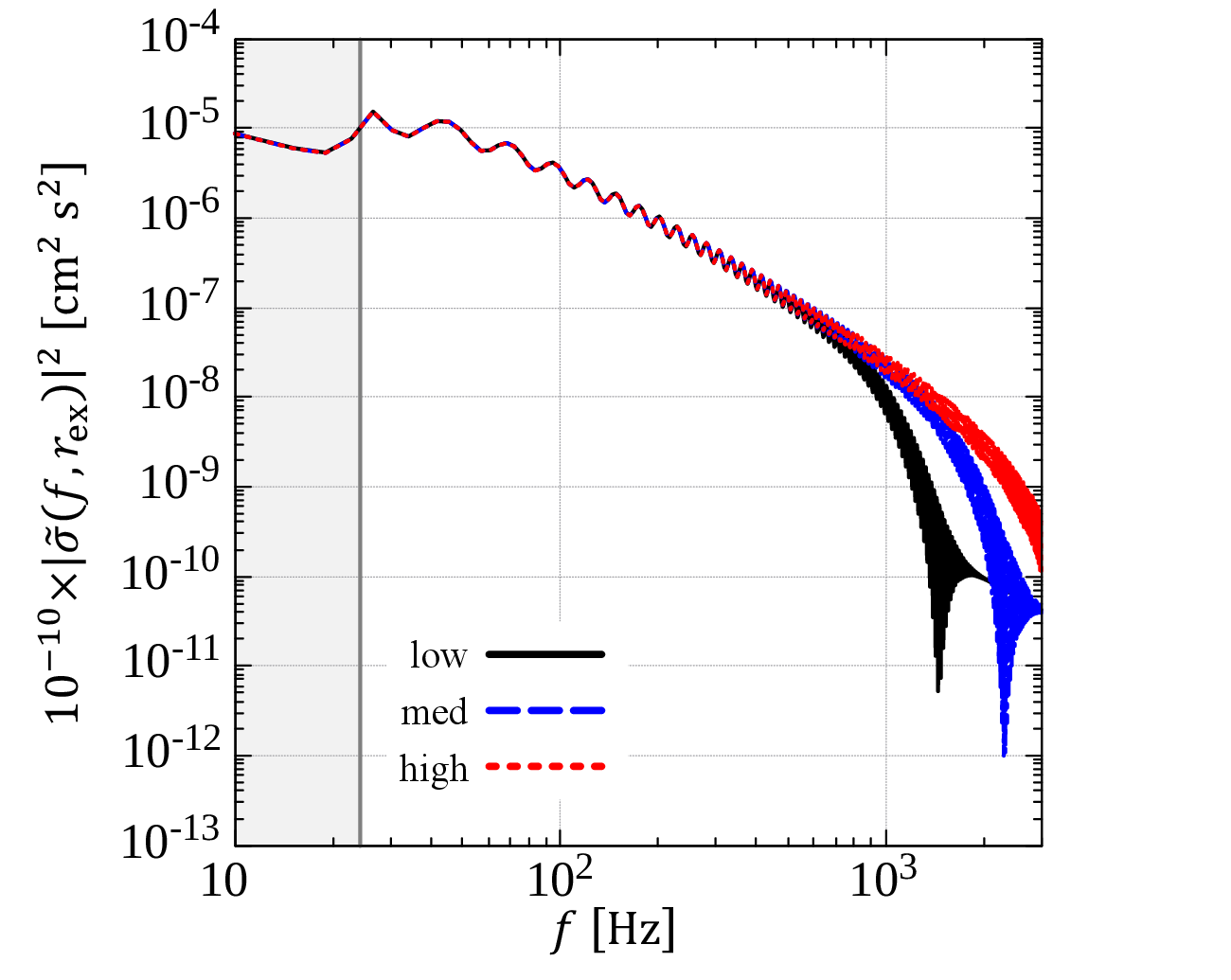}
    \caption{
    Extracted spectra with $\Delta r_\mathrm{low}$, $\Delta r_\mathrm{med}$, and $\Delta r_\mathrm{high}$.
    The gray-filed area $(f<\omega_{*}/2\pi=24.2~\mathrm{Hz})$ shows nonpropagating modes.
    }
    \label{Fig:extract_resolution}
  \end{minipage}
  \hspace{0.05\hsize}
  \begin{minipage}[t]{0.45\hsize}
    \centering
    \includegraphics
    [clip,keepaspectratio,width=8cm]
    {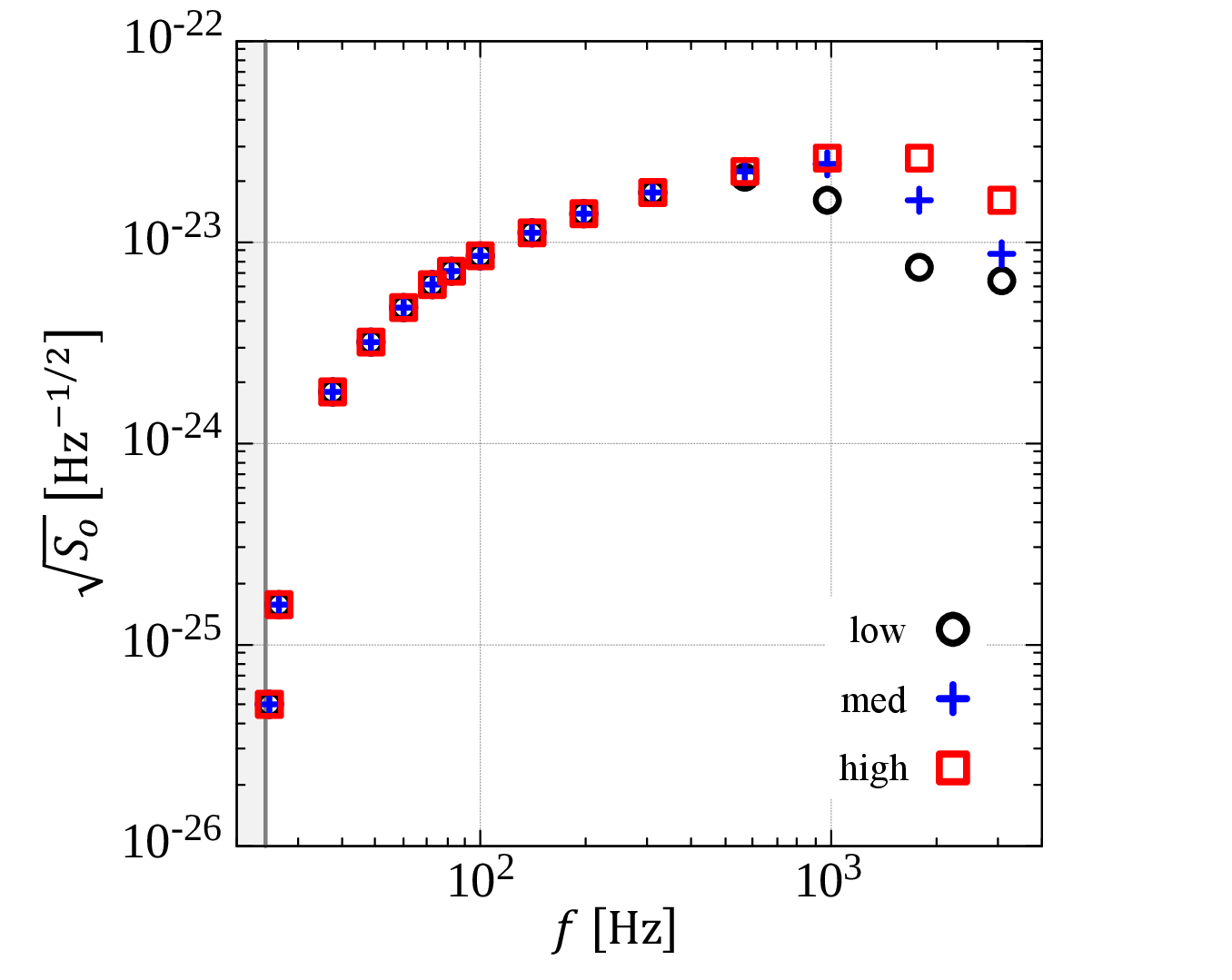}
    \caption{
    Observed spectra with $\Delta r_\mathrm{low}$, $\Delta r_\mathrm{med}$, and $\Delta r_\mathrm{high}$.
    The gray-filed area $(f<\omega_{*}/2\pi=24.2~\mathrm{Hz})$ shows the no scalar GW signals area.
    }
    \label{Fig:observed_resolution}
  \end{minipage}
\end{figure*}

\subsection{Shock tube}
The relativistic shock tube problem is the special relativistic hydrodynamics test problem.
Thus, we fix the metric as the Minkowski spacetime at all times,
\begin{eqnarray}
  \alpha=1
  ~,~~~
  X=1
  ~,
\end{eqnarray}
and we adopt the Cartesian coordinate $0\le x\le1$ with the uniform grid spacing $\Delta x=10^{-3}$.
We use the $\Gamma$-law type EOS with $\Gamma_\mathrm{th}=5/3$.
Following~\cite{marti2003numerical},
we set the initial profile as shown in Table~\ref{Table:STP_initial}.
Note that
we set the pressure of the right side $(x>0.5)$ to be the nonzero value,
but the small value,
unlike Ref.~\cite{marti2003numerical}.

Figure \ref{Fig:STP} shows profiles of the density, pressure, and velocity at $t=0.4$.
As can be seen from Fig.~\ref{Fig:STP}, the numerical solutions agree well with the exact solutions,
which demonstrates the capability of our code in special-relativistic hydrodynamics.

\subsection{Oppenheimer-Snyder collapse}
The Oppenheimer-Snyder collapse is the general relativistic test problem,
in which pressureless dust collapses to a black hole in spherical symmetry.
We set the initial condition as below following~\cite{o2010new}.
Dust with the total baryon mass of $M=1M_{\odot}$ is uniformly distributed in the sphere
of which the radius is $R=10M$.
In the outside of this dust sphere,
we set the atmosphere density $\rho_\mathrm{atmo}=1~\mathrm{g/cm^3}$.
We use the polytropic EOS with $K=10^{-20}$ and $\Gamma=5/3$~\cite{o2010new}
to set the small pressure.

We use the coordinate in which the line element is given by Eq.~(\ref{Eq:Line-Element}) with $F=1$,
and we set the radial coordinate $0\le r\le20M$ with the uniform grid spacing $\Delta r=2.5\times10^{-3}M$.
Figure~\ref{Fig:OSC} displays snap shots of the density (left panel) and $\alpha$ (right panel).
The numerical solutions show good agreement with the exact solutions,
which shows the validity of our code in the general relativistic collapse in spherical symmetry.

\section{CONVERGENCE CHECK}\label{Sec_Convergence}
In this appendix,
we show the convergence test of our code.
We set parameters to be $\mu=10^{-13}~\mathrm{eV}$, $\alpha_{0}=10^{-2}$, and $\beta_{0}=-20$
and use the cubic self-interaction potential with $\lambda_{3}=10^{3}$ in this convergence test.

To check the convergence of our code,
we performed three core collapse simulations with different uniform grid resolutions inside $r=40~\mathrm{km}$:
$\Delta r_\mathrm{low}=100~\mathrm{m}$, $\Delta r_\mathrm{med}=50~\mathrm{m}$, and $\Delta r_\mathrm{high}=25~\mathrm{m}$.
Different total grid points are used for each grid resolution:
$N_\mathrm{low}=5000$, $N_\mathrm{med}=10000$, and $N_\mathrm{high}=20000$, respectively.

The convergence factor $Q$ of a quantity $q$ is defined by
\begin{eqnarray}
  Q
  :=
  \frac{q_\mathrm{low}-q_\mathrm{med}}
  {q_\mathrm{med}-q_\mathrm{high}}
  =
  \frac{(\Delta r_\mathrm{low})^n-(\Delta r_\mathrm{med})^n}
  {(\Delta r_\mathrm{med})^n-(\Delta r_\mathrm{high})^n}
  ~,
\end{eqnarray}
where $n$ is the accuracy of the numerical calculation.
For our grid condition,
$Q=2$ and $Q=4$ are expected for the first-order and second-order convergence, respectively.
Figure~\ref{Fig:convergence} shows differences between extracted scalar waveforms with different grid resolutions.
From Fig.~\ref{Fig:convergence},
numerical solutions converge between the first and second order, which we expected.

For another diagnosis,
we also checked the conservation of the total baryon mass in the computation domain as shown in Fig.~\ref{Fig:mass}.
From Fig.~\ref{Fig:mass},
the total baryon mass is conserved by about $10^{-2}\%$ for all grid resolutions,
and the finer grid resolution improves the conservation of the total baryon mass.

Finally, we checked the dependence of scalar GW signals on the grid resolution.
As shown in Fig.~\ref{Fig:extract_resolution}, the extracted spectra damp in the higher frequency region
$(f\gtrsim10^{3}~\mathrm{Hz})$ due to a lack of grid resolution.
In the lower frequency region, on the other hand, the GW spectra converge well.
From Figs.~\ref{Fig:extract_resolution} and \ref{Fig:observed_resolution}, in particular,
the slope of the extracted and observed spectra does not change.
Therefore, we believe that our discussion to distinguish the self-interaction of the scalar field based on
the power-law index of the declined observed spectra is not affected by the grid resolution.

\section{DEPENDENCE OF THE POWER-LAW INDEX ON THE EOS AND PROGENITOR}\label{Sec_EOS_Progenitor}

In this appendix, we check the dependence of scalar GWs on the progenitor and EOS paying particular attention to the power-law index of the spectra.
Detailed studies for the dependence of the dynamics of the collapse and emission of GWs on the EOS will be found in Refs.~\cite{rosca2020core,cheong2019numerical}.
Here, we set parameters to be $\mu=10^{-13}~\mathrm{eV}$, $\alpha_{0}=10^{-2}$, $\beta_{0}=-20$,
$\lambda_{3}=10^{6}$ for the cubic self-interaction, and $\lambda_{4}=10^{8}$ for the quartic self-interaction.

\begin{figure}[t!]
  \includegraphics
  [clip,keepaspectratio,width=8cm]
  {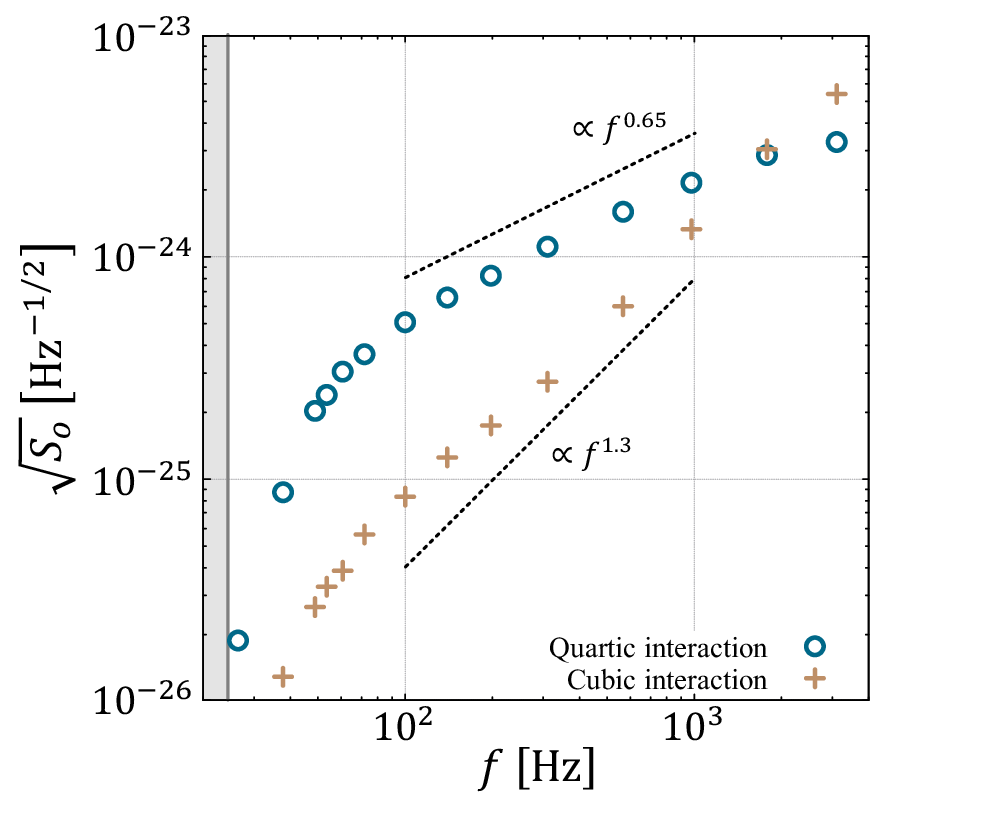}
  \caption{
  Observed spectra for the generalized piecewise polytropic EOS.
  The gray-filed area $(f<\omega_{*}/2\pi=24.2~\mathrm{Hz})$ shows the no scalar GW signals area.
  }
  \label{Fig:EOS}
\end{figure}

\subsection{Core collapse simulations with a hybrid EOS with a continuous speed of sound}

In this paper, following the previous studies~\cite{rosca2019inverse,sperhake2017long,rosca2020core,cheong2019numerical,huang2021scalar,gerosa2016numerical},
we adopt the cold part of the hybrid EOS as the piecewise polytropic EOS Eq.~(\ref{Eq:Cold_Part}),
in which the sound speed is discontinuous at $\rho_\mathrm{nuc}$.
Thus, we investigate the effects of the discontinuous speed of sound on the power-law index of the declined observed spectra.
Note that the dependence of the emission of GWs on the parameters $(\Gamma_{1},\Gamma_{2},\Gamma_\mathrm{th})$
has been studied in Refs.~\cite{rosca2020core,cheong2019numerical}.
It is found that the GW spectra do not strongly depend on the parameters.

\begin{figure}[t!]
  \includegraphics
  [clip,keepaspectratio,width=8cm]
  {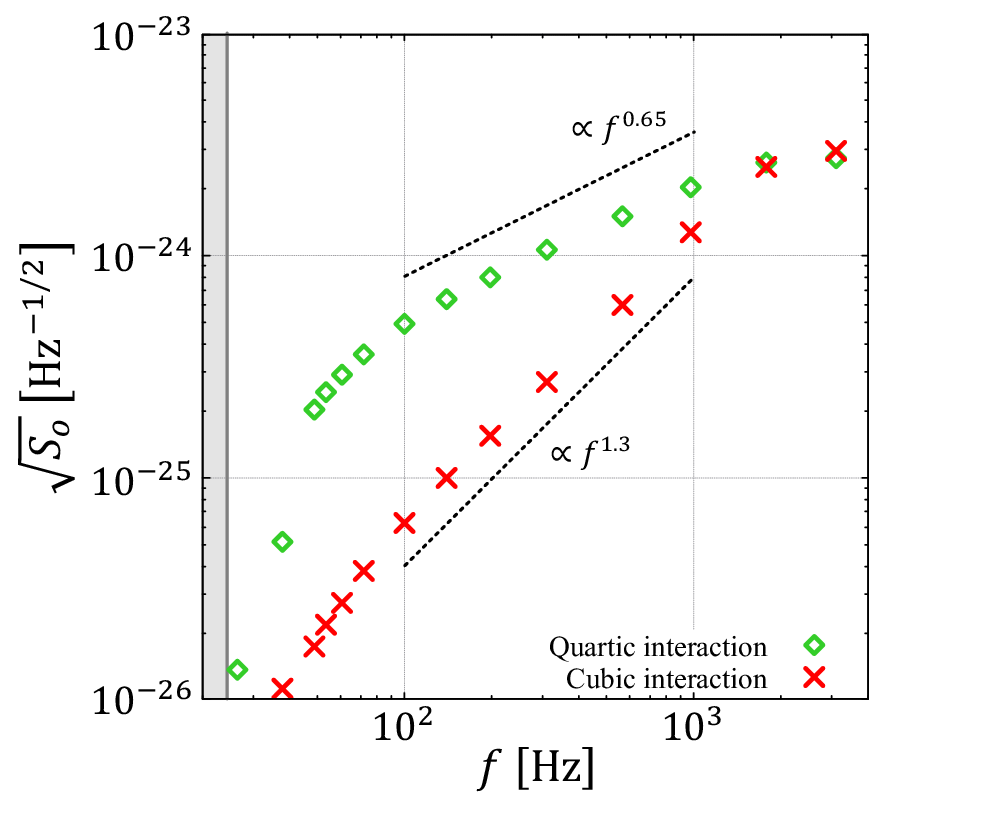}
  \caption{
  Observed spectra for the progenitor model with  mass of $40M_{\odot}$.
  The gray-filed area $(f<\omega_{*}/2\pi=24.2~\mathrm{Hz})$ shows the no scalar GW signals area.
  }
  \label{Fig:progenitor}
\end{figure}

The piecewise polytropic EOS with a continuous speed of sound, so-called generalized piecewise polytropic EOS, is given by~\cite{PhysRevD.102.083027}
\begin{eqnarray}
  \label{Eq:generalized_piecewise}
  P_\mathrm{cold}
  &=&
  \left\{\begin{array}{lr}
  \displaystyle
  K_{1}\rho^{\Gamma_{1}}
  &
  \rho\le\rho_\mathrm{nuc}
  \\
  \displaystyle
  \tilde{K}_{2}\rho^{\Gamma_{2}}+\Lambda
  &
  \rho>\rho_\mathrm{nuc}
  \end{array}\right.
  ~,
  \\
  \epsilon_\mathrm{cold}
  &=&
  \left\{\begin{array}{lc}
  \displaystyle
  \frac{K_{1}\rho^{\Gamma_{1}-1}}{\Gamma_{1}-1}
  &
  \rho\le\rho_\mathrm{nuc}
  \\\\
  \displaystyle
  \frac{\tilde{K}_{2}\rho^{\Gamma_{2}-1}}{\Gamma_{2}-1}
  -\frac{\Lambda}{\rho}\\
  \displaystyle~~~~~
  +\frac{\Gamma_{1}K_{1}\rho_\mathrm{nuc}^{\Gamma_{1}-1}}{\Gamma_{1}-1}
  -\frac{\Gamma_{2}\tilde{K}_{2}\rho_\mathrm{nuc}^{\Gamma_{2}-1}}{\Gamma_{2}-1}
  &
  \rho>\rho_\mathrm{nuc}
  \end{array}\right.
  ~.
  \nonumber\\
\end{eqnarray}
Here, the additional constant term $\Lambda$ is given as follows~\cite{PhysRevD.102.083027}
\begin{eqnarray}
  \Lambda
  &=&
  \biggr(
  1-\frac{\Gamma_{1}}{\Gamma_{2}}
  \biggr)
  K_{1}\rho^{\Gamma_{1}}_\mathrm{nuc}
  ~,
\end{eqnarray}
and makes the sound speed to be continuous at $\rho_\mathrm{nuc}$.
Note that in the generalized piecewise polytropic EOS,
the coefficient $\tilde{K}_{2}$ is determined from the differentiability of the pressure at $\rho_\mathrm{nuc}$,
so $\tilde{K}_{2}\neq K_{2}$.
On the other hand, $\Lambda$ is determined from the continuity of the pressure at $\rho_\mathrm{nuc}$.
We set the parameters of the EOS to be the same values as the main text of this paper:
$(\Gamma_{1},\Gamma_{2},\Gamma_\mathrm{th})=(1.3,2.5,1.35)$ and $K_{1}=4.9345\times10^{14}~\mathrm{cgs}$.

Figure~\ref{Fig:EOS} shows observed spectra for the hybrid EOS consisting of the generalized piecewise polytropic EOS Eq.~(\ref{Eq:generalized_piecewise}) and the thermal part Eq.~(\ref{Eq:Thermal_Part}).
From Fig.~\ref{Fig:EOS},
the power-law index of the declined observed spectra is unchanged.

\begin{figure*}[t!]
  \includegraphics
  [clip,keepaspectratio,width=16cm]
  {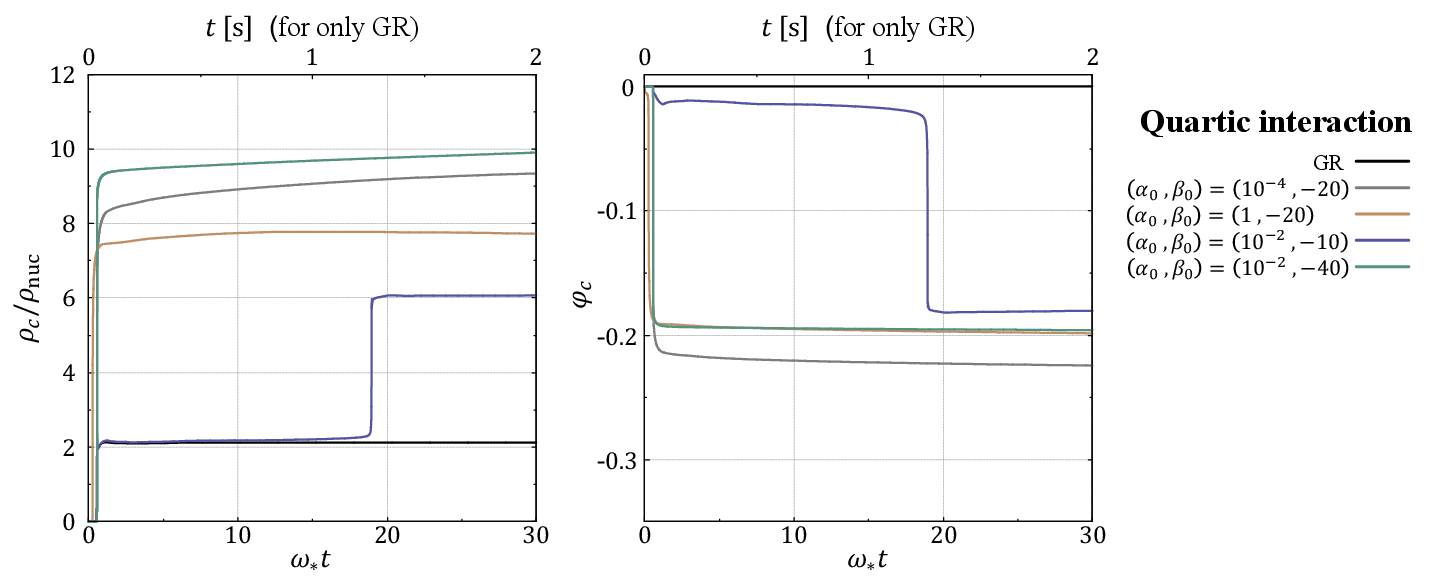}
  \caption{
  Time evolution of
  the central density $\rho_{c}:=\rho(t,r=0)$ (left panel)
  and the central scalar field $\varphi_{c}:=\varphi(t,r=0)$ (right panel)
  for the quartic self-interaction.
  The scalar-field mass and self-interaction parameter are fixed as $\mu=10^{-14}~\mathrm{eV}$ and $\lambda_{4}=10^{8}$, while $(\alpha_{0},\beta_{0})$ are varied.
  The lower horizontal axis is the normalized time $\omega_{*}t$,
  and upper horizontal axis is the time $t$ for only the GR model.
  The model with $(\alpha_{0},\beta_{0})=(10^{-2},-10)$ is the two-stage collapse model,
  and the second core bounce time of this model is $\omega_{*}t_{b,2}\approx19$.
  }
  \label{Fig:dynamics_conformal}
\end{figure*}
\begin{figure*}[!]
  \includegraphics
  [clip,keepaspectratio,width=16cm]
  {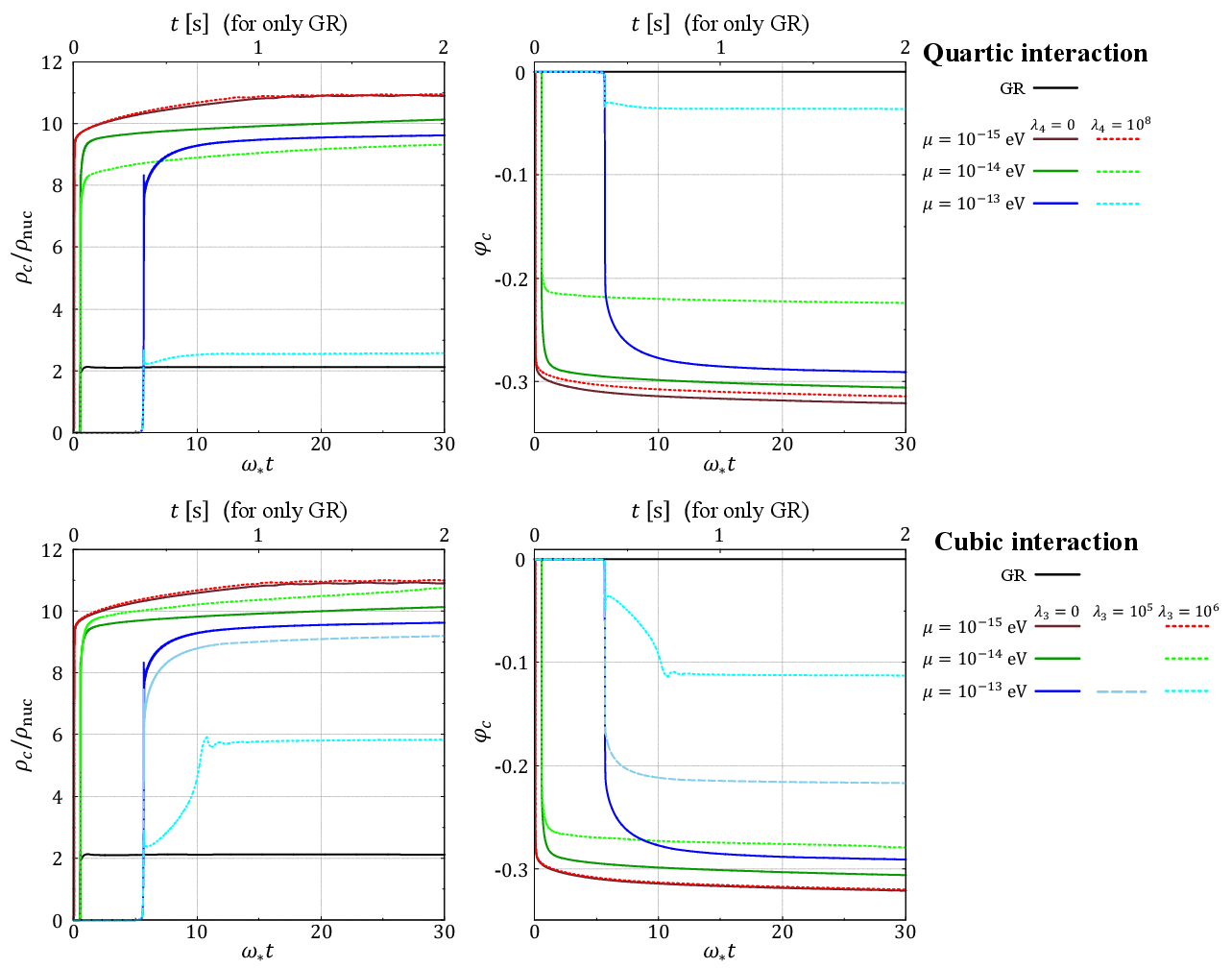}
  \caption{
  Similar to Fig.~\ref{Fig:dynamics_conformal},
  but for the models with fixed $(\alpha_{0},\beta_{0})=(10^{-2},-20)$,
  while $\mu$ and $\lambda_{i}$ are varied.
  Top and bottom panels are for the quartic self-interaction and cubic self-interaction, respectively.
  Note that for the cubic self-interaction model with $\mu=10^{-13}~\mathrm{eV}$,
  $\rho_{c}$ and $\varphi_{c}$ for the model with $\lambda_{3}=10^{5}$ are also shown.
  }
  \label{Fig:dynamics_mass}
\end{figure*}

\subsection{Dependence on the progenitor}

We also investigate the dependence of the power-law index of the GW spectra on the progenitor.
Here, we adopt a massive progenitor model with mass of $40M_{\odot}$~\cite{woosley2002evolution} (the
model s40 in Ref.~\cite{rosca2020core}).
Figure~\ref{Fig:progenitor} shows observed GW spectra for this progenitor model.
It is found that the power-law index of the spectra is unchanged.
This is likely to because the structure of the core does not strongly depend on the progenitor mass (see Ref.~\cite{rosca2020core} for a more detailed study of the dependence of collapse dynamics and scalar GWs on the progenitor in the case of the massive ST theory without self-interactions).

\section{COLLAPSE DYNAMICS}\label{Sec_Collapse-dynamics}
In this appendix, we briefly summarize the dynamics of the collapse.

\subsection{Overall features of the dynamics of the collapse}

Figure~\ref{Fig:dynamics_conformal} shows time evolution of the central density (left panel) and the central value of the scalar field (right panel) as functions of the normalized time $\omega_{*}t$ for the quartic self-interaction.
The scalar-field mass and self-interaction parameter are fixed as
$\mu=10^{-14}~\mathrm{eV}$ and $\lambda_{4}=10^{8}$, while $(\alpha_{0},\beta_{0})$ are varied.
Except for the model with $(\alpha_{0},\beta_{0})=(10^{-2},-10)$, a scalarized neutron star is formed promptly after the core bounce.
The core bounce time is $\omega_{*}t_{b}\approx0.56$
for the models with $(\alpha_{0},\beta_{0})=(10^{-4},-20),(10^{-2},-40)$
and $\omega_{*}t_{b}\approx0.26$ for the model with $(\alpha_{0},\beta_{0})=(1,-20)$.
This is the so-called single-stage collapse~\cite{rosca2020core,cheong2019numerical,huang2021scalar}.
For the model with $(\alpha_{0},\beta_{0})=(10^{-2},-10)$, on the other hand, the core experiences two bounces before the strongly scalarized neutron star is formed.
In this model,
the core collapses to a weakly scalarized neutron star after the first bounce at $\omega_{*}t_{b}\approx0.56$,
and then it subsequently collapses to a strongly scalarized neutron star after the second bounce
at $\omega_{*}t_{b,2}\approx19$.
This is the so-called multistage collapse found in the previous studies~\cite{rosca2020core,huang2021scalar}.

\subsection{Influence of the self-interaction on the dynamics of the collapse}

As shown in the previous studies, we found that the scalarization of a neutron star is suppressed by the self-interaction of the scalar field.
Figure~\ref{Fig:dynamics_mass} shows time evolution of the central density and the central scalar field for the quartic self-interaction (top panels) and cubic self-interaction (bottom panels) for different values of $\mu$ and $\lambda_{i}$.
The values of $(\alpha_{0},\beta_{0})$ are fixed as $(10^{-2},-20)$.
As can be seen in Fig.~\ref{Fig:dynamics_mass},
the self-interaction suppresses the scalarization of a neutron star
(compare models with $\lambda_{i}=0$ and $\lambda_{i}\neq0$) as found in the previous
studies~\cite{cheong2019numerical,huang2021scalar}.

We also found that the suppression of the scalarization is more prominent for the more massive scalar field.
In addition, we found a complicated scalarization process for the cubic self-interaction model with $\mu=10^{-13}~\mathrm{eV}$ and $\lambda_{3}=10^{6}$, thanks to the comprehensive parameter study performed in this paper.
This complicated scalarization process is likely to be explained as follows.
In this model, the prompt collapse to a strongly scalarized neutron star is prevented due to the self-interaction, and the core collapses to a weakly scalarized neutron star at the bounce at $\omega_{*}t_{b}\approx5.6$.
At the bounce, the released gravitational energy is smaller than that of the prompt collapse to a strongly scalarized neutron star, and a weaker shock is formed.
As a consequence, the accretion onto the core continues and pushes back the shock,
triggering the subsequent scalarization of the neutron star at $\omega_{*}t\approx10.5$.
This result indicates that when $\lambda_{3}$ is above some critical value, the self-interaction not
only suppresses the scalarization of a neutron star, but can also affect time evolution of the scalar field.
\footnote{The previous study~\cite{huang2021scalar} shows that the self-interaction eliminates one intermediate stage from the multi-stage collapse to a black hole when $\lambda_{3}$ reaches the critical value.}

The dynamics of the collapse depend on the scalar-field mass,
self-interaction parameters, and the conformal factor.
However, this information is weakly imprinted in observed scalar GW signals.
This is likely due to the inverse chirp structure of the scalar GW signals.

\nocite{*}

\bibliography{./main.bbl}

\begin{thebibliography}{45}%
\makeatletter
\providecommand \@ifxundefined [1]{%
 \@ifx{#1\undefined}
}%
\providecommand \@ifnum [1]{%
 \ifnum #1\expandafter \@firstoftwo
 \else \expandafter \@secondoftwo
 \fi
}%
\providecommand \@ifx [1]{%
 \ifx #1\expandafter \@firstoftwo
 \else \expandafter \@secondoftwo
 \fi
}%
\providecommand \natexlab [1]{#1}%
\providecommand \enquote  [1]{``#1''}%
\providecommand \bibnamefont  [1]{#1}%
\providecommand \bibfnamefont [1]{#1}%
\providecommand \citenamefont [1]{#1}%
\providecommand \href@noop [0]{\@secondoftwo}%
\providecommand \href [0]{\begingroup \@sanitize@url \@href}%
\providecommand \@href[1]{\@@startlink{#1}\@@href}%
\providecommand \@@href[1]{\endgroup#1\@@endlink}%
\providecommand \@sanitize@url [0]{\catcode `\\12\catcode `\$12\catcode
  `\&12\catcode `\#12\catcode `\^12\catcode `\_12\catcode `\%12\relax}%
\providecommand \@@startlink[1]{}%
\providecommand \@@endlink[0]{}%
\providecommand \url  [0]{\begingroup\@sanitize@url \@url }%
\providecommand \@url [1]{\endgroup\@href {#1}{\urlprefix }}%
\providecommand \urlprefix  [0]{URL }%
\providecommand \Eprint [0]{\href }%
\providecommand \doibase [0]{https://doi.org/}%
\providecommand \selectlanguage [0]{\@gobble}%
\providecommand \bibinfo  [0]{\@secondoftwo}%
\providecommand \bibfield  [0]{\@secondoftwo}%
\providecommand \translation [1]{[#1]}%
\providecommand \BibitemOpen [0]{}%
\providecommand \bibitemStop [0]{}%
\providecommand \bibitemNoStop [0]{.\EOS\space}%
\providecommand \EOS [0]{\spacefactor3000\relax}%
\providecommand \BibitemShut  [1]{\csname bibitem#1\endcsname}%
\let\auto@bib@innerbib\@empty
\bibitem [{\citenamefont {Psaltis}(2008)}]{psaltis2008probes}%
  \BibitemOpen
  \bibfield  {author} {\bibinfo {author} {\bibfnamefont {D.}~\bibnamefont
  {Psaltis}},\ }\href@noop {} {\bibfield  {journal} {\bibinfo  {journal}
  {Living Rev. Relativity}\ }\textbf {\bibinfo {volume} {11}},\ \bibinfo
  {pages} {9} (\bibinfo {year} {2008})}\BibitemShut {NoStop}%
\bibitem [{\citenamefont {Will}(2014)}]{will2014confrontation}%
  \BibitemOpen
  \bibfield  {author} {\bibinfo {author} {\bibfnamefont {C.~M.}\ \bibnamefont
  {Will}},\ }\href@noop {} {\bibfield  {journal} {\bibinfo  {journal} {Living
  Rev. Relativity}\ }\textbf {\bibinfo {volume} {17}},\ \bibinfo {pages} {4}
  (\bibinfo {year} {2014})}\BibitemShut {NoStop}%
\bibitem [{\citenamefont {Berti}\ \emph {et~al.}(2015)\citenamefont {Berti}
  \emph {et~al.}}]{berti2015testing}%
  \BibitemOpen
  \bibfield  {author} {\bibinfo {author} {\bibfnamefont {E.}~\bibnamefont
  {Berti}} \emph {et~al.},\ }\href@noop {} {\bibfield  {journal} {\bibinfo
  {journal} {Classical Quantum Gravity}\ }\textbf {\bibinfo {volume} {32}},\
  \bibinfo {pages} {243001} (\bibinfo {year} {2015})}\BibitemShut {NoStop}%
\bibitem [{\citenamefont {Li}\ \emph {et~al.}(2011)\citenamefont {Li},
  \citenamefont {Li}, \citenamefont {Wang},\ and\ \citenamefont
  {Wang}}]{li2011dark}%
  \BibitemOpen
  \bibfield  {author} {\bibinfo {author} {\bibfnamefont {M.}~\bibnamefont
  {Li}}, \bibinfo {author} {\bibfnamefont {X.-D.}\ \bibnamefont {Li}}, \bibinfo
  {author} {\bibfnamefont {S.}~\bibnamefont {Wang}},\ and\ \bibinfo {author}
  {\bibfnamefont {Y.}~\bibnamefont {Wang}},\ }\href@noop {} {\bibfield
  {journal} {\bibinfo  {journal} {Commun. Theor. Phys.}\ }\textbf {\bibinfo
  {volume} {56}},\ \bibinfo {pages} {525} (\bibinfo {year} {2011})}\BibitemShut
  {NoStop}%
\bibitem [{\citenamefont {Burgess}(2004)}]{burgess2004quantum}%
  \BibitemOpen
  \bibfield  {author} {\bibinfo {author} {\bibfnamefont {C.~P.}\ \bibnamefont
  {Burgess}},\ }\href@noop {} {\bibfield  {journal} {\bibinfo  {journal}
  {Living Rev. Relativity}\ }\textbf {\bibinfo {volume} {7}},\ \bibinfo {pages}
  {5} (\bibinfo {year} {2004})}\BibitemShut {NoStop}%
\bibitem [{\citenamefont {Damour}\ and\ \citenamefont
  {Esposito-Farese}(1992)}]{damour1992tensor}%
  \BibitemOpen
  \bibfield  {author} {\bibinfo {author} {\bibfnamefont {T.}~\bibnamefont
  {Damour}}\ and\ \bibinfo {author} {\bibfnamefont {G.}~\bibnamefont
  {Esposito-Farese}},\ }\href@noop {} {\bibfield  {journal} {\bibinfo
  {journal} {Classical Quantum Gravity}\ }\textbf {\bibinfo {volume} {9}},\
  \bibinfo {pages} {2093} (\bibinfo {year} {1992})}\BibitemShut {NoStop}%
\bibitem [{\citenamefont {Fujii}\ and\ \citenamefont
  {Maeda}(2003)}]{fujii2003scalar}%
  \BibitemOpen
  \bibfield  {author} {\bibinfo {author} {\bibfnamefont {Y.}~\bibnamefont
  {Fujii}}\ and\ \bibinfo {author} {\bibfnamefont {K.}~\bibnamefont {Maeda}},\
  }\href@noop {} {\emph {\bibinfo {title} {The Scalar-Tensor Theory of
  Gravitation}}}\ (\bibinfo  {publisher} {Cambridge University Press,
  Cambridge, England},\ \bibinfo {year} {2003})\BibitemShut {NoStop}%
\bibitem [{\citenamefont {La}\ and\ \citenamefont
  {Steinhardt}(1989)}]{la1989extended}%
  \BibitemOpen
  \bibfield  {author} {\bibinfo {author} {\bibfnamefont {D.}~\bibnamefont
  {La}}\ and\ \bibinfo {author} {\bibfnamefont {P.~J.}\ \bibnamefont
  {Steinhardt}},\ }\href@noop {} {\bibfield  {journal} {\bibinfo  {journal}
  {Phys. Rev. Lett.}\ }\textbf {\bibinfo {volume} {62}},\ \bibinfo {pages}
  {376} (\bibinfo {year} {1989})}\BibitemShut {NoStop}%
\bibitem [{\citenamefont {Sotiriou}\ and\ \citenamefont
  {Faraoni}(2010)}]{sotiriou2010f}%
  \BibitemOpen
  \bibfield  {author} {\bibinfo {author} {\bibfnamefont {T.~P.}\ \bibnamefont
  {Sotiriou}}\ and\ \bibinfo {author} {\bibfnamefont {V.}~\bibnamefont
  {Faraoni}},\ }\href@noop {} {\bibfield  {journal} {\bibinfo  {journal} {Rev.
  Mod. Phys.}\ }\textbf {\bibinfo {volume} {82}},\ \bibinfo {pages} {451}
  (\bibinfo {year} {2010})}\BibitemShut {NoStop}%
\bibitem [{\citenamefont {Nojiri}\ and\ \citenamefont
  {Odintsov}(2003)}]{nojiri2003modified}%
  \BibitemOpen
  \bibfield  {author} {\bibinfo {author} {\bibfnamefont {S.}~\bibnamefont
  {Nojiri}}\ and\ \bibinfo {author} {\bibfnamefont {S.~D.}\ \bibnamefont
  {Odintsov}},\ }\href@noop {} {\bibfield  {journal} {\bibinfo  {journal}
  {Phys. Rev. D}\ }\textbf {\bibinfo {volume} {68}},\ \bibinfo {pages} {123512}
  (\bibinfo {year} {2003})}\BibitemShut {NoStop}%
\bibitem [{\citenamefont {Callan}\ \emph {et~al.}(1985)\citenamefont {Callan},
  \citenamefont {Friedan}, \citenamefont {Martinec},\ and\ \citenamefont
  {Perry}}]{callan1985strings}%
  \BibitemOpen
  \bibfield  {author} {\bibinfo {author} {\bibfnamefont {C.~G.}\ \bibnamefont
  {Callan}}, \bibinfo {author} {\bibfnamefont {D.}~\bibnamefont {Friedan}},
  \bibinfo {author} {\bibfnamefont {E.}~\bibnamefont {Martinec}},\ and\
  \bibinfo {author} {\bibfnamefont {M.}~\bibnamefont {Perry}},\ }\href@noop {}
  {\bibfield  {journal} {\bibinfo  {journal} {Nucl. Phys.}\ }\textbf {\bibinfo
  {volume} {B262}},\ \bibinfo {pages} {593} (\bibinfo {year}
  {1985})}\BibitemShut {NoStop}%
\bibitem [{\citenamefont {Damour}\ and\ \citenamefont
  {Esposito-Farese}(1993)}]{damour1993nonperturbative}%
  \BibitemOpen
  \bibfield  {author} {\bibinfo {author} {\bibfnamefont {T.}~\bibnamefont
  {Damour}}\ and\ \bibinfo {author} {\bibfnamefont {G.}~\bibnamefont
  {Esposito-Farese}},\ }\href@noop {} {\bibfield  {journal} {\bibinfo
  {journal} {Phys. Rev. Lett.}\ }\textbf {\bibinfo {volume} {70}},\ \bibinfo
  {pages} {2220} (\bibinfo {year} {1993})}\BibitemShut {NoStop}%
\bibitem [{\citenamefont {Damour}\ and\ \citenamefont
  {Esposito-Farese}(1996)}]{damour1996tensor}%
  \BibitemOpen
  \bibfield  {author} {\bibinfo {author} {\bibfnamefont {T.}~\bibnamefont
  {Damour}}\ and\ \bibinfo {author} {\bibfnamefont {G.}~\bibnamefont
  {Esposito-Farese}},\ }\href@noop {} {\bibfield  {journal} {\bibinfo
  {journal} {Phys. Rev. D}\ }\textbf {\bibinfo {volume} {54}},\ \bibinfo
  {pages} {1474} (\bibinfo {year} {1996})}\BibitemShut {NoStop}%
\bibitem [{\citenamefont {Ramazano{\u{g}}lu}\ and\ \citenamefont
  {Pretorius}(2016)}]{ramazanouglu2016spontaneous}%
  \BibitemOpen
  \bibfield  {author} {\bibinfo {author} {\bibfnamefont {F.~M.}\ \bibnamefont
  {Ramazano{\u{g}}lu}}\ and\ \bibinfo {author} {\bibfnamefont {F.}~\bibnamefont
  {Pretorius}},\ }\href@noop {} {\bibfield  {journal} {\bibinfo  {journal}
  {Phys. Rev. D}\ }\textbf {\bibinfo {volume} {93}},\ \bibinfo {pages} {064005}
  (\bibinfo {year} {2016})}\BibitemShut {NoStop}%
\bibitem [{\citenamefont {Morisaki}\ and\ \citenamefont
  {Suyama}(2017)}]{morisaki2017spontaneous}%
  \BibitemOpen
  \bibfield  {author} {\bibinfo {author} {\bibfnamefont {S.}~\bibnamefont
  {Morisaki}}\ and\ \bibinfo {author} {\bibfnamefont {T.}~\bibnamefont
  {Suyama}},\ }\href@noop {} {\bibfield  {journal} {\bibinfo  {journal} {Phys.
  Rev. D}\ }\textbf {\bibinfo {volume} {96}},\ \bibinfo {pages} {084026}
  (\bibinfo {year} {2017})}\BibitemShut {NoStop}%
\bibitem [{\citenamefont {Rosca-Mead}\ \emph
  {et~al.}(2020{\natexlab{a}})\citenamefont {Rosca-Mead}, \citenamefont
  {Moore}, \citenamefont {Sperhake}, \citenamefont {Agathos},\ and\
  \citenamefont {Gerosa}}]{rosca2020structure}%
  \BibitemOpen
  \bibfield  {author} {\bibinfo {author} {\bibfnamefont {R.}~\bibnamefont
  {Rosca-Mead}}, \bibinfo {author} {\bibfnamefont {C.~J.}\ \bibnamefont
  {Moore}}, \bibinfo {author} {\bibfnamefont {U.}~\bibnamefont {Sperhake}},
  \bibinfo {author} {\bibfnamefont {M.}~\bibnamefont {Agathos}},\ and\ \bibinfo
  {author} {\bibfnamefont {D.}~\bibnamefont {Gerosa}},\ }\href@noop {}
  {\bibfield  {journal} {\bibinfo  {journal} {Symmetry}\ }\textbf {\bibinfo
  {volume} {12}},\ \bibinfo {pages} {1384} (\bibinfo {year}
  {2020}{\natexlab{a}})}\BibitemShut {NoStop}%
\bibitem [{\citenamefont {Staykov}\ \emph {et~al.}(2018)\citenamefont
  {Staykov}, \citenamefont {Popchev}, \citenamefont {Doneva},\ and\
  \citenamefont {Yazadjiev}}]{staykov2018static}%
  \BibitemOpen
  \bibfield  {author} {\bibinfo {author} {\bibfnamefont {K.~V.}\ \bibnamefont
  {Staykov}}, \bibinfo {author} {\bibfnamefont {D.}~\bibnamefont {Popchev}},
  \bibinfo {author} {\bibfnamefont {D.~D.}\ \bibnamefont {Doneva}},\ and\
  \bibinfo {author} {\bibfnamefont {S.~S.}\ \bibnamefont {Yazadjiev}},\
  }\href@noop {} {\bibfield  {journal} {\bibinfo  {journal} {Eur. Phys. J. C}\
  }\textbf {\bibinfo {volume} {78}},\ \bibinfo {pages} {586} (\bibinfo {year}
  {2018})}\BibitemShut {NoStop}%
\bibitem [{\citenamefont {Yazadjiev}\ \emph {et~al.}(2016)\citenamefont
  {Yazadjiev}, \citenamefont {Doneva},\ and\ \citenamefont
  {Popchev}}]{yazadjiev2016slowly}%
  \BibitemOpen
  \bibfield  {author} {\bibinfo {author} {\bibfnamefont {S.~S.}\ \bibnamefont
  {Yazadjiev}}, \bibinfo {author} {\bibfnamefont {D.~D.}\ \bibnamefont
  {Doneva}},\ and\ \bibinfo {author} {\bibfnamefont {D.}~\bibnamefont
  {Popchev}},\ }\href@noop {} {\bibfield  {journal} {\bibinfo  {journal} {Phys.
  Rev. D}\ }\textbf {\bibinfo {volume} {93}},\ \bibinfo {pages} {084038}
  (\bibinfo {year} {2016})}\BibitemShut {NoStop}%
\bibitem [{\citenamefont {Doneva}\ \emph {et~al.}(2013)\citenamefont {Doneva},
  \citenamefont {Yazadjiev}, \citenamefont {Stergioulas},\ and\ \citenamefont
  {Kokkotas}}]{doneva2013rapidly}%
  \BibitemOpen
  \bibfield  {author} {\bibinfo {author} {\bibfnamefont {D.~D.}\ \bibnamefont
  {Doneva}}, \bibinfo {author} {\bibfnamefont {S.~S.}\ \bibnamefont
  {Yazadjiev}}, \bibinfo {author} {\bibfnamefont {N.}~\bibnamefont
  {Stergioulas}},\ and\ \bibinfo {author} {\bibfnamefont {K.~D.}\ \bibnamefont
  {Kokkotas}},\ }\href@noop {} {\bibfield  {journal} {\bibinfo  {journal}
  {Phys. Rev. D}\ }\textbf {\bibinfo {volume} {88}},\ \bibinfo {pages} {084060}
  (\bibinfo {year} {2013})}\BibitemShut {NoStop}%
\bibitem [{\citenamefont {Kuan}\ \emph {et~al.}(2022)\citenamefont {Kuan},
  \citenamefont {Suvorov}, \citenamefont {Doneva},\ and\ \citenamefont
  {Yazadjiev}}]{PhysRevLett.129.121104}%
  \BibitemOpen
  \bibfield  {author} {\bibinfo {author} {\bibfnamefont {H.-J.}\ \bibnamefont
  {Kuan}}, \bibinfo {author} {\bibfnamefont {A.~G.}\ \bibnamefont {Suvorov}},
  \bibinfo {author} {\bibfnamefont {D.~D.}\ \bibnamefont {Doneva}},\ and\
  \bibinfo {author} {\bibfnamefont {S.~S.}\ \bibnamefont {Yazadjiev}},\
  }\href@noop {} {\bibfield  {journal} {\bibinfo  {journal} {Phys. Rev. Lett.}\
  }\textbf {\bibinfo {volume} {129}},\ \bibinfo {pages} {121104} (\bibinfo
  {year} {2022})}\BibitemShut {NoStop}%
\bibitem [{\citenamefont {Bertotti}\ \emph {et~al.}(2003)\citenamefont
  {Bertotti}, \citenamefont {Iess},\ and\ \citenamefont
  {Tortora}}]{bertotti2003test}%
  \BibitemOpen
  \bibfield  {author} {\bibinfo {author} {\bibfnamefont {B.}~\bibnamefont
  {Bertotti}}, \bibinfo {author} {\bibfnamefont {L.}~\bibnamefont {Iess}},\
  and\ \bibinfo {author} {\bibfnamefont {P.}~\bibnamefont {Tortora}},\
  }\href@noop {} {\bibfield  {journal} {\bibinfo  {journal} {Nature (London)}\
  }\textbf {\bibinfo {volume} {425}},\ \bibinfo {pages} {374} (\bibinfo {year}
  {2003})}\BibitemShut {NoStop}%
\bibitem [{\citenamefont {Antoniadis}\ \emph {et~al.}(2013)\citenamefont
  {Antoniadis} \emph {et~al.}}]{antoniadis2013massive}%
  \BibitemOpen
  \bibfield  {author} {\bibinfo {author} {\bibfnamefont {J.}~\bibnamefont
  {Antoniadis}} \emph {et~al.},\ }\href@noop {} {\bibfield  {journal} {\bibinfo
   {journal} {Science}\ }\textbf {\bibinfo {volume} {340}},\ \bibinfo {pages}
  {1233232} (\bibinfo {year} {2013})}\BibitemShut {NoStop}%
\bibitem [{\citenamefont {Freire}\ \emph {et~al.}(2012)\citenamefont {Freire}
  \emph {et~al.}}]{freire2012relativistic}%
  \BibitemOpen
  \bibfield  {author} {\bibinfo {author} {\bibfnamefont {P.~C.~C.}\
  \bibnamefont {Freire}} \emph {et~al.},\ }\href@noop {} {\bibfield  {journal}
  {\bibinfo  {journal} {Mon. Not. R. Astron. Soc.}\ }\textbf {\bibinfo {volume}
  {423}},\ \bibinfo {pages} {3328} (\bibinfo {year} {2012})}\BibitemShut
  {NoStop}%
\bibitem [{\citenamefont {Novak}(1998)}]{novak1998neutron}%
  \BibitemOpen
  \bibfield  {author} {\bibinfo {author} {\bibfnamefont {J.}~\bibnamefont
  {Novak}},\ }\href@noop {} {\bibfield  {journal} {\bibinfo  {journal} {Phys.
  Rev. D}\ }\textbf {\bibinfo {volume} {58}},\ \bibinfo {pages} {064019}
  (\bibinfo {year} {1998})}\BibitemShut {NoStop}%
\bibitem [{\citenamefont {Chatziioannou}\ \emph {et~al.}(2012)\citenamefont
  {Chatziioannou}, \citenamefont {Yunes},\ and\ \citenamefont
  {Cornish}}]{chatziioannou2012model}%
  \BibitemOpen
  \bibfield  {author} {\bibinfo {author} {\bibfnamefont {K.}~\bibnamefont
  {Chatziioannou}}, \bibinfo {author} {\bibfnamefont {N.}~\bibnamefont
  {Yunes}},\ and\ \bibinfo {author} {\bibfnamefont {N.}~\bibnamefont
  {Cornish}},\ }\href@noop {} {\bibfield  {journal} {\bibinfo  {journal} {Phys.
  Rev. D}\ }\textbf {\bibinfo {volume} {86}},\ \bibinfo {pages} {022004}
  (\bibinfo {year} {2012})}\BibitemShut {NoStop}%
\bibitem [{\citenamefont {Aasi}\ \emph {et~al.}(2015)\citenamefont {Aasi} \emph
  {et~al.}}]{aasi2015advanced}%
  \BibitemOpen
  \bibfield  {author} {\bibinfo {author} {\bibfnamefont {J.}~\bibnamefont
  {Aasi}} \emph {et~al.},\ }\href@noop {} {\bibfield  {journal} {\bibinfo
  {journal} {Classical Quantum Gravity}\ }\textbf {\bibinfo {volume} {32}},\
  \bibinfo {pages} {074001} (\bibinfo {year} {2015})}\BibitemShut {NoStop}%
\bibitem [{\citenamefont {Acernese}\ \emph {et~al.}(2014)\citenamefont
  {Acernese} \emph {et~al.}}]{acernese2014advanced}%
  \BibitemOpen
  \bibfield  {author} {\bibinfo {author} {\bibfnamefont {F.}~\bibnamefont
  {Acernese}} \emph {et~al.},\ }\href@noop {} {\bibfield  {journal} {\bibinfo
  {journal} {Classical Quantum Gravity}\ }\textbf {\bibinfo {volume} {32}},\
  \bibinfo {pages} {024001} (\bibinfo {year} {2014})}\BibitemShut {NoStop}%
\bibitem [{\citenamefont {Somiya}(2012)}]{somiya2012detector}%
  \BibitemOpen
  \bibfield  {author} {\bibinfo {author} {\bibfnamefont {K.}~\bibnamefont
  {Somiya}},\ }\href@noop {} {\bibfield  {journal} {\bibinfo  {journal}
  {Classical Quantum Gravity}\ }\textbf {\bibinfo {volume} {29}},\ \bibinfo
  {pages} {124007} (\bibinfo {year} {2012})}\BibitemShut {NoStop}%
\bibitem [{\citenamefont {Abbott}\ \emph {et~al.}(2020)\citenamefont {Abbott}
  \emph {et~al.}}]{abbott2020prospects}%
  \BibitemOpen
  \bibfield  {author} {\bibinfo {author} {\bibfnamefont {B.~P.}\ \bibnamefont
  {Abbott}} \emph {et~al.},\ }\href@noop {} {\bibfield  {journal} {\bibinfo
  {journal} {Living Rev. Relativity}\ }\textbf {\bibinfo {volume} {23}},\
  \bibinfo {pages} {3} (\bibinfo {year} {2020})}\BibitemShut {NoStop}%
\bibitem [{\citenamefont {Punturo}\ \emph {et~al.}(2010)\citenamefont {Punturo}
  \emph {et~al.}}]{punturo2010einstein}%
  \BibitemOpen
  \bibfield  {author} {\bibinfo {author} {\bibfnamefont {M.}~\bibnamefont
  {Punturo}} \emph {et~al.},\ }\href@noop {} {\bibfield  {journal} {\bibinfo
  {journal} {Classical Quantum Gravity}\ }\textbf {\bibinfo {volume} {27}},\
  \bibinfo {pages} {194002} (\bibinfo {year} {2010})}\BibitemShut {NoStop}%
\bibitem [{\citenamefont {Abbott}\ \emph {et~al.}(2017)\citenamefont {Abbott}
  \emph {et~al.}}]{abbott2017exploring}%
  \BibitemOpen
  \bibfield  {author} {\bibinfo {author} {\bibfnamefont {B.~P.}\ \bibnamefont
  {Abbott}} \emph {et~al.},\ }\href@noop {} {\bibfield  {journal} {\bibinfo
  {journal} {Classical Quantum Gravity}\ }\textbf {\bibinfo {volume} {34}},\
  \bibinfo {pages} {044001} (\bibinfo {year} {2017})}\BibitemShut {NoStop}%
\bibitem [{\citenamefont {Rosca-Mead}\ \emph
  {et~al.}(2020{\natexlab{b}})\citenamefont {Rosca-Mead}, \citenamefont
  {Sperhake}, \citenamefont {Moore}, \citenamefont {Agathos}, \citenamefont
  {Gerosa},\ and\ \citenamefont {Ott}}]{rosca2020core}%
  \BibitemOpen
  \bibfield  {author} {\bibinfo {author} {\bibfnamefont {R.}~\bibnamefont
  {Rosca-Mead}}, \bibinfo {author} {\bibfnamefont {U.}~\bibnamefont
  {Sperhake}}, \bibinfo {author} {\bibfnamefont {C.~J.}\ \bibnamefont {Moore}},
  \bibinfo {author} {\bibfnamefont {M.}~\bibnamefont {Agathos}}, \bibinfo
  {author} {\bibfnamefont {D.}~\bibnamefont {Gerosa}},\ and\ \bibinfo {author}
  {\bibfnamefont {C.~D.}\ \bibnamefont {Ott}},\ }\href@noop {} {\bibfield
  {journal} {\bibinfo  {journal} {Phys. Rev. D}\ }\textbf {\bibinfo {volume}
  {102}},\ \bibinfo {pages} {044010} (\bibinfo {year}
  {2020}{\natexlab{b}})}\BibitemShut {NoStop}%
\bibitem [{\citenamefont {Sperhake}\ \emph {et~al.}(2017)\citenamefont
  {Sperhake}, \citenamefont {Moore}, \citenamefont {Rosca}, \citenamefont
  {Agathos}, \citenamefont {Gerosa},\ and\ \citenamefont
  {Ott}}]{sperhake2017long}%
  \BibitemOpen
  \bibfield  {author} {\bibinfo {author} {\bibfnamefont {U.}~\bibnamefont
  {Sperhake}}, \bibinfo {author} {\bibfnamefont {C.~J.}\ \bibnamefont {Moore}},
  \bibinfo {author} {\bibfnamefont {R.}~\bibnamefont {Rosca}}, \bibinfo
  {author} {\bibfnamefont {M.}~\bibnamefont {Agathos}}, \bibinfo {author}
  {\bibfnamefont {D.}~\bibnamefont {Gerosa}},\ and\ \bibinfo {author}
  {\bibfnamefont {C.~D.}\ \bibnamefont {Ott}},\ }\href@noop {} {\bibfield
  {journal} {\bibinfo  {journal} {Phys. Rev. Lett.}\ }\textbf {\bibinfo
  {volume} {119}},\ \bibinfo {pages} {201103} (\bibinfo {year}
  {2017})}\BibitemShut {NoStop}%
\bibitem [{\citenamefont {Rosca-Mead}\ \emph {et~al.}(2019)\citenamefont
  {Rosca-Mead}, \citenamefont {Moore}, \citenamefont {Agathos},\ and\
  \citenamefont {Sperhake}}]{rosca2019inverse}%
  \BibitemOpen
  \bibfield  {author} {\bibinfo {author} {\bibfnamefont {R.}~\bibnamefont
  {Rosca-Mead}}, \bibinfo {author} {\bibfnamefont {C.~J.}\ \bibnamefont
  {Moore}}, \bibinfo {author} {\bibfnamefont {M.}~\bibnamefont {Agathos}},\
  and\ \bibinfo {author} {\bibfnamefont {U.}~\bibnamefont {Sperhake}},\
  }\href@noop {} {\bibfield  {journal} {\bibinfo  {journal} {Classical Quantum
  Gravity}\ }\textbf {\bibinfo {volume} {36}},\ \bibinfo {pages} {134003}
  (\bibinfo {year} {2019})}\BibitemShut {NoStop}%
\bibitem [{\citenamefont {Cheong}\ and\ \citenamefont
  {Li}(2019)}]{cheong2019numerical}%
  \BibitemOpen
  \bibfield  {author} {\bibinfo {author} {\bibfnamefont {P.~C.-K.}\
  \bibnamefont {Cheong}}\ and\ \bibinfo {author} {\bibfnamefont {T.~G.~F.}\
  \bibnamefont {Li}},\ }\href@noop {} {\bibfield  {journal} {\bibinfo
  {journal} {Phys. Rev. D}\ }\textbf {\bibinfo {volume} {100}},\ \bibinfo
  {pages} {024027} (\bibinfo {year} {2019})}\BibitemShut {NoStop}%
\bibitem [{\citenamefont {Huang}\ \emph {et~al.}(2021)\citenamefont {Huang},
  \citenamefont {Geng},\ and\ \citenamefont {Kuan}}]{huang2021scalar}%
  \BibitemOpen
  \bibfield  {author} {\bibinfo {author} {\bibfnamefont {D.}~\bibnamefont
  {Huang}}, \bibinfo {author} {\bibfnamefont {C.-Q.}\ \bibnamefont {Geng}},\
  and\ \bibinfo {author} {\bibfnamefont {H.-J.}\ \bibnamefont {Kuan}},\
  }\href@noop {} {\bibfield  {journal} {\bibinfo  {journal} {Classical Quantum
  Gravity}\ }\textbf {\bibinfo {volume} {38}},\ \bibinfo {pages} {245006}
  (\bibinfo {year} {2021})}\BibitemShut {NoStop}%
\bibitem [{\citenamefont {Gerosa}\ \emph {et~al.}(2016)\citenamefont {Gerosa},
  \citenamefont {Sperhake},\ and\ \citenamefont {Ott}}]{gerosa2016numerical}%
  \BibitemOpen
  \bibfield  {author} {\bibinfo {author} {\bibfnamefont {D.}~\bibnamefont
  {Gerosa}}, \bibinfo {author} {\bibfnamefont {U.}~\bibnamefont {Sperhake}},\
  and\ \bibinfo {author} {\bibfnamefont {C.~D.}\ \bibnamefont {Ott}},\
  }\href@noop {} {\bibfield  {journal} {\bibinfo  {journal} {Classical Quantum
  Gravity}\ }\textbf {\bibinfo {volume} {33}},\ \bibinfo {pages} {135002}
  (\bibinfo {year} {2016})}\BibitemShut {NoStop}%
\bibitem [{\citenamefont {Salgado}(2006)}]{salgado2006cauchy}%
  \BibitemOpen
  \bibfield  {author} {\bibinfo {author} {\bibfnamefont {M.}~\bibnamefont
  {Salgado}},\ }\href@noop {} {\bibfield  {journal} {\bibinfo  {journal}
  {Classical Quantum Gravity}\ }\textbf {\bibinfo {volume} {23}},\ \bibinfo
  {pages} {4719} (\bibinfo {year} {2006})}\BibitemShut {NoStop}%
\bibitem [{\citenamefont {Janka}\ \emph {et~al.}(1993)\citenamefont {Janka},
  \citenamefont {Zwerger},\ and\ \citenamefont {Moenchmeyer}}]{janka1993does}%
  \BibitemOpen
  \bibfield  {author} {\bibinfo {author} {\bibfnamefont {H.-T.}\ \bibnamefont
  {Janka}}, \bibinfo {author} {\bibfnamefont {T.}~\bibnamefont {Zwerger}},\
  and\ \bibinfo {author} {\bibfnamefont {R.}~\bibnamefont {Moenchmeyer}},\
  }\href@noop {} {\bibfield  {journal} {\bibinfo  {journal} {Astron.
  Astrophys.}\ }\textbf {\bibinfo {volume} {268}},\ \bibinfo {pages} {360}
  (\bibinfo {year} {1993})}\BibitemShut {NoStop}%
\bibitem [{\citenamefont {Dimmelmeier}\ \emph {et~al.}(2002)\citenamefont
  {Dimmelmeier}, \citenamefont {Font},\ and\ \citenamefont
  {M{\"u}ller}}]{dimmelmeier2002relativistic}%
  \BibitemOpen
  \bibfield  {author} {\bibinfo {author} {\bibfnamefont {H.}~\bibnamefont
  {Dimmelmeier}}, \bibinfo {author} {\bibfnamefont {J.~A.}\ \bibnamefont
  {Font}},\ and\ \bibinfo {author} {\bibfnamefont {E.}~\bibnamefont
  {M{\"u}ller}},\ }\href@noop {} {\bibfield  {journal} {\bibinfo  {journal}
  {Astron. Astrophys.}\ }\textbf {\bibinfo {volume} {388}},\ \bibinfo {pages}
  {917} (\bibinfo {year} {2002})}\BibitemShut {NoStop}%
\bibitem [{\citenamefont {Shapiro}\ and\ \citenamefont
  {Teukolsky}(1983)}]{shapiro2008black}%
  \BibitemOpen
  \bibfield  {author} {\bibinfo {author} {\bibfnamefont {S.~L.}\ \bibnamefont
  {Shapiro}}\ and\ \bibinfo {author} {\bibfnamefont {S.~A.}\ \bibnamefont
  {Teukolsky}},\ }\href@noop {} {\emph {\bibinfo {title} {Black Holes, White
  Dwarfs, and Neutron Stars: The Physics of Compact Objects}}}\ (\bibinfo
  {publisher} {John Wiley \& Sons, New York},\ \bibinfo {year}
  {1983})\BibitemShut {NoStop}%
\bibitem [{\citenamefont {Woosley}\ \emph {et~al.}(2002)\citenamefont
  {Woosley}, \citenamefont {Heger},\ and\ \citenamefont
  {Weaver}}]{woosley2002evolution}%
  \BibitemOpen
  \bibfield  {author} {\bibinfo {author} {\bibfnamefont {S.~E.}\ \bibnamefont
  {Woosley}}, \bibinfo {author} {\bibfnamefont {A.}~\bibnamefont {Heger}},\
  and\ \bibinfo {author} {\bibfnamefont {T.~A.}\ \bibnamefont {Weaver}},\
  }\href@noop {} {\bibfield  {journal} {\bibinfo  {journal} {Rev. Mod. Phys.}\
  }\textbf {\bibinfo {volume} {74}},\ \bibinfo {pages} {1015} (\bibinfo {year}
  {2002})}\BibitemShut {NoStop}%
\bibitem [{\citenamefont {Mart{\'\i}}\ and\ \citenamefont
  {M{\"u}ller}(2003)}]{marti2003numerical}%
  \BibitemOpen
  \bibfield  {author} {\bibinfo {author} {\bibfnamefont {J.~M.}\ \bibnamefont
  {Mart{\'\i}}}\ and\ \bibinfo {author} {\bibfnamefont {E.}~\bibnamefont
  {M{\"u}ller}},\ }\href@noop {} {\bibfield  {journal} {\bibinfo  {journal}
  {Living Rev. Relativity}\ }\textbf {\bibinfo {volume} {6}},\ \bibinfo {pages}
  {7} (\bibinfo {year} {2003})}\BibitemShut {NoStop}%
\bibitem [{\citenamefont {O'Connor}\ and\ \citenamefont
  {Ott}(2010)}]{o2010new}%
  \BibitemOpen
  \bibfield  {author} {\bibinfo {author} {\bibfnamefont {E.}~\bibnamefont
  {O'Connor}}\ and\ \bibinfo {author} {\bibfnamefont {C.~D.}\ \bibnamefont
  {Ott}},\ }\href@noop {} {\bibfield  {journal} {\bibinfo  {journal} {Classical
  Quantum Gravity}\ }\textbf {\bibinfo {volume} {27}},\ \bibinfo {pages}
  {114103} (\bibinfo {year} {2010})}\BibitemShut {NoStop}%
\bibitem [{\citenamefont {O'Boyle}\ \emph {et~al.}(2020)\citenamefont
  {O'Boyle}, \citenamefont {Markakis}, \citenamefont {Stergioulas},\ and\
  \citenamefont {Read}}]{PhysRevD.102.083027}%
  \BibitemOpen
  \bibfield  {author} {\bibinfo {author} {\bibfnamefont {M.~F.}\ \bibnamefont
  {O'Boyle}}, \bibinfo {author} {\bibfnamefont {C.}~\bibnamefont {Markakis}},
  \bibinfo {author} {\bibfnamefont {N.}~\bibnamefont {Stergioulas}},\ and\
  \bibinfo {author} {\bibfnamefont {J.~S.}\ \bibnamefont {Read}},\ }\href@noop
  {} {\bibfield  {journal} {\bibinfo  {journal} {Phys. Rev. D}\ }\textbf
  {\bibinfo {volume} {102}},\ \bibinfo {pages} {083027} (\bibinfo {year}
  {2020})}\BibitemShut {NoStop}%
\end{thebibliography}%

\end{document}